\documentclass[12pt]{article}
\usepackage[latin9]{inputenc}
\usepackage{geometry}
\usepackage{float}
\usepackage{amsmath}
\usepackage{amsthm}
\usepackage{amssymb}
\usepackage{stmaryrd}
\usepackage{graphicx}
\usepackage{setspace}
\usepackage[authoryear]{natbib}
\usepackage[unicode=true,
 bookmarks=false,
 breaklinks=false,pdfborder={0 0 1},backref=section,colorlinks=false]
 {hyperref}
\hypersetup{
 colorlinks,citecolor=blue,pdftex}

\makeatletter

\providecommand{\tabularnewline}{\\}

\usepackage{amsfonts}
\usepackage{amsthm}
\usepackage{lscape}
\usepackage{appendix}
\usepackage{dcolumn}
\usepackage{rotating}
\usepackage{comment}
\usepackage{color}

\usepackage{diagbox}
\usepackage[flushleft]{threeparttable}
\usepackage{subfigure}
\usepackage{pgf}
\usepackage{tikz}\usepackage{caption}
\usepackage{tkz-tab}
\usepackage{tikz-3dplot}
\usetikzlibrary{calc}
\usetikzlibrary{arrows, automata}

\usepackage{changepage}
\usepackage{booktabs}

\newcolumntype{d}[1]{D{.}{.}{#1}}
\newcolumntype{t}[1]{D{,}{,}{#1}}
\newcolumntype{i}[1]{D{.}{}{#1}}
\newtheorem{theorem}{Theorem}[section]

\newtheorem{corollary}{Corollary}[section]

\newtheorem{definition}{Definition}[section]
\newtheorem{example}{Example}
\newtheorem{lemma}{Lemma}[section]

\newtheorem{proposition}{Proposition}[section]

\theoremstyle{plain}

\newtheorem{innercustomas}{Assumption}\newenvironment{myas}[1]{\renewcommand\theinnercustomas{#1}\innercustomas}{\endinnercustomas}

\numberwithin{equation}{section}

\makeatother

\begin{document}
\title{Policy Learning with Distributional Welfare\thanks{For helpful discussions, the authors are grateful to Debopam Bhattacharya,
Kei Hirano, Nathan Kallus, Toru Kitagawa, Ashesh Rambachan, Vira Semenova, and participants
at the 2024 Econometric Society Interdisciplinary Frontiers (ESIF)
Conference, the 2024 International Society for Clinical Biostatistics
Conference, the Advances in Econometrics Conference 2023 and the seminar
participants at Brown and Essex. We also thank Xuanman Li for her
excellent research assistance. Yifan Cui acknowledges the financial
support from the National Key R\&D Program of China (2024YFA1015600), and the National Natural Science Foundation of China (12471266 and U23A2064).}}
\author{Yifan Cui\\
 Center for Data Science\\
 Zhejiang University\\
 \UrlFont\href{mailto:cuiyf\%5C\%5C\%5C\%40zju.edu.cn}{cuiyf@zju.edu.cn}\and
Sukjin Han\\
 School of Economics\\
 University of Bristol\\
 \UrlFont\href{mailto:vincent.han\%5C\%5C\%5C\%40bristol.ac.uk}{vincent.han@bristol.ac.uk}}
\date{\today}

\maketitle
\vspace{-0.2cm}

\begin{abstract}
In this paper, we explore optimal treatment allocation policies that
target distributional welfare. Most literature on treatment choice
has considered utilitarian welfare based on the conditional average
treatment effect (ATE). While average welfare is intuitive, it may
yield undesirable allocations especially when individuals are heterogeneous
(e.g., with outliers)---the very reason individualized treatments
were introduced in the first place. This observation motivates us
to propose an optimal policy that allocates the treatment based on
the conditional \emph{quantile of individual treatment effects} (QoTE).
Depending on the choice of the quantile probability, this criterion
can accommodate a policymaker who is either prudent or negligent.
The challenge of identifying the QoTE lies in its requirement for
knowledge of the joint distribution of the counterfactual outcomes,
which is not generally point-identified. We introduce minimax policies
that are robust to this model uncertainty. A range of identifying
assumptions can be used to yield more informative policies. For both
stochastic and deterministic policies, we establish the asymptotic
bound on the regret of implementing the proposed policies. The framework
can be generalized to any setting where welfare is defined as a functional
of the joint distribution of the potential outcomes.

\vspace{0.1in}


\end{abstract}

\section{Introduction\label{sec:Introduction}}

Individuals are heterogeneous, so are their responses to treatments
or programs. When designing policies (e.g., rules of allocating treatments
or programs), it is important to reflect the heterogeneity of individual
treatment effects. A policymaker (PM), or equivalently an analyst,
would devise a policy to achieve a specific objective (e.g., welfare).
Depending on how the PM aggregates individual gains, her objective
can be viewed as either \emph{utilitarian} or \emph{non-utilitarian}.
A utilitarian PM would consider welfare that takes the sum or average
of individual gains to ensure the greatest benefits for the greatest
number, whereas a non-utilitarian (e.g., prioritarian, maximin) PM
would prioritize specific groups of individuals. The utilitarian objective
has been the most widely-used criterion in the literature of treatment
allocations and policy learning (e.g., \citet{manski2004statistical};
see below for a further review). However, there may be settings where
the utilitarian goal is less sensible. For example, the target population
may exhibit skewed heterogeneity (e.g., outliers). As another example,
the PM may want to target a vulnerable population or privileged individuals,
or a certain share of benefited individuals.\footnote{The possibility of non-utilitarian welfare is also briefly mentioned
in \citet{manski2004statistical}.} The purpose of this paper is to explore objectives of a (non-utilitarian)
PM who is concerned with certain aspects of the distribution (e.g.,
tails) of treatment effects or who has political incentives and thus
makes decisions influenced by vote shares.

In this paper, we develop a policy learning framework that concerns
distributional welfare. A policy is defined as a mapping from individuals'
observed characteristics to either a deterministic or stochastic decision
of treatment allocation. Intuitively, the knowledge of individual
treatment effects conditional on characteristics plays a crucial role
in learning such a policy. We propose an objective function that is
formulated based on the conditional quantile of individual treatment
effects (QoTE). This objective function is robust to outliers of treatment
effects and, more importantly, can reflect the PM's level of prudence
toward the target population. As quantifying the uncertainty of allocation
decisions is intrinsically difficult (e.g., \citet{chen2023inference}),
the ability to adjust the level of prudence can be practically valuable
to the PM.

Suppose the PM employs the utilitarian welfare, which can be written
as a function of the conditional average treatment effect (ATE). If
the policy class is unconstrained, it is optimal for the utilitarian
PM to treat each subgroup (defined by observed characteristics) whenever
their ATE is positive. Suppose that this PM faces a target subgroup,
say black females, whose distribution of treatment effects exhibits
that a small share of individuals enjoys positive treatment effects
that dominate the negative effects of the remaining majority. If the
resulting ATE is positive, then the PM would treat \emph{all} black
females, harming the majority. The objective function based on the
QoTE with the quantile probability $\tau=0.5$ (i.e., the median of
treatment effects) would not suffer from this sensitivity to outliers.
Moreover, the PM can choose the quantile probability $\tau$ (i.e.,
the rank in individual treatment effects) to set a reference group.
A large $\tau$ corresponds to a PM who is willing to focus on privileged
individuals in each subgroup, ignoring the majority of less advantaged,
thus being a \emph{negligent} PM. A small $\tau$ corresponds to a
PM who is concerned with the disadvantaged, treating each subgroup
only if most benefit from the treatment, thus being a \emph{prudent}
PM. Relatedly, we show that the PM equipped with the QoTE can be interpreted
as being concerned with vote shares when each individual casts a vote
whenever he or she experiences a positive gain from the treatment.

An alternative objective function that can be robust to certain outliers
is the one based on the conditional quantile treatment effect (QTE)
which contrasts the quantiles of treated and untreated outcomes. We
argue that this quantity may not be a desirable basis for individualized
treatment decisions, because an individual represented by the quantile
of treated outcomes is not necessarily the same individual represented
by the same quantile of untreated outcomes. For example, as shown
below, it is difficult for the PM to aim the level of prudence when
the criterion is based on the QTE. On the other hand, the QoTE by
definition captures an individual with a specific rank in gains and
thus naturally accommodates the notion of prudence of PM.

Despite the desirable properties of the PM's objective function constructed
from the QoTE, the challenge is that the QoTE is not generally point-identified
even when the PM has access to experimental data. This is due to the
fact that the joint distribution of counterfactual outcomes is involved
in the definition of the QoTE. We therefore propose a minimax criterion
that is robust to model ambiguity. In particular, we propose to minimize
the worst-case regret calculated over the class of joint distributions
of counterfactual outcomes that are compatible with the data and identifying
assumptions. We then show that a range of identifying assumptions
that can be imposed to tighten the identified set of the QoTE, sometimes
to a singleton, leading to more informative policies. These assumptions
can be imposed by practitioners depending on their specific settings.
For some assumptions, bounds on the QoTE may not have a closed-form
expression. In this case, an optimization algorithm can be used to
compute the bounds. By using a Bernstein approximation, we show how
the optimization problem becomes a simple linear programming.

We establish theoretical properties of the proposed minimax policy
by providing asymptotic bounds on the regret of implementing the estimated
policy. First, when the policy class is unconstrained, we show that
the estimated policy is consistent if either the bounds on the QoTE
are sign-determining or the QoTE is point-identified. Otherwise, the
leading term of the regret bound has a magnitude that depends on the
relative location of zero in the QoTE bounds. We provide the theory
for both stochastic and deterministic policies. The leading term with
the stochastic policy is smaller than that with the deterministic
policy, consistent with the findings in the literature \citep{manski2007minimax,stoye2007minimax,cui2021individualized}.
It is important to allow the policy class to be constrained as the
PM may prefer a parsimonious policy or face institutional or budget
constraints. In this case of constrained policy classes, we propose
to use the machine learning (ML) technique of the outcome-weighting
framework with a surrogate loss \citep{zhao2012estimating}. We
then show that the ML-estimated policy is consistent and characterize
the rate in terms of approximation and estimation errors.

In this paper, we consider empirical applications in two well-known
randomized control trials in medicine and economics. The first application
concerns the allocation of a diagnostic procedure for critically ill
patients using data from the Study to Understand Prognoses and Preferences
for Outcomes and Risks of Treatments \citep{hirano2001estimation}.
The second application examines the allocation of job training using
data from the US National Job Training Partnership Act \citep{bloom1997benefits}.
In both applications, a common finding is that there exists substantial
heterogeneity in the distributional treatment effects and thus in
the corresponding allocation decisions based on the QoTE. To deliver
the main messages of this paper, we show in the space of covariates
how the allocation decisions take place (see Figures \ref{fig:application1-2}--\ref{fig:application1-3}
below). As expected, the allocation becomes more aggressive as the
quantile probability $\tau$ increases. We compare this result with
the decisions based on the QTE and ATE. The QTE decisions do not exhibit
the change in the degree of prudence in $\tau$. Comparing the ATE
decisions with the QoTE decisions with $\tau=0.5$, we can inspect
whether outliers are problematic in calculating the ATE decision in
these data sets. In this sense, we view the QoTE decisions as a means
of a robustness check for the ATE decisions prevalent in the literature.

The policy learning framework of this paper can be generalized to
any setting where welfare is defined as a functional of the joint
distribution of potential outcomes. Towards the end of the paper,
we introduce a general framework and propose other examples of welfare
criteria that may be interest a non-utilitarian PM. These include
criteria targeting individuals who are either worst off in the counterfactual
baseline or worst-affected by the treatment.

\subsection{Related Literature}

Learning optimal treatment regimes has received considerable interest
in the past few years across multiple disciplines including computer
science \citep{dudik2011doubly}, econometrics \citep{manski2004statistical,hirano2009asymptotics,stoye2009minimax,kitagawa2018should,athey2021policy,mbakop2021model,ida2022choosing},
and statistics \citep{murphy2003optimal,kosorok2015adaptive,kosorok2019precision,tsiatis2019dynamic,jiang2019entropy}.
In statistics, existing methods for learning optimal treatment regimes
are mostly through either Q-learning \citep{watkins1992q,qian2011performance}
or A-learning \citep{murphy2003optimal,robins2004optimal,shi2018high}.
Alternative approaches have emerged from a classification perspective
\citep{zhao2012estimating,zhang2012robust,rubin2012statistical},
which has proven more robust to model misspecification in some settings.

Recently, there is a growing literature on learning optimal treatment
allocations that aims to relax the unconfoundedness assumption. Within
this literature, a strand of work considers cases where the welfare
and optimal treatment regime is point-identified, that is, the treatment
decision is free from ambiguity given the observed data. \citet{cui2021semiparametric,qiu2021optimal}
consider instrumental variable (IV) approaches under a point identification
and \citet{han2021comment,cui2021necessary} consider IV methods under
a sign identification. \citet{kallus2021causal,qi2023proximal,shen2023optimal}
consider optimal policy learning under the proximal causal inference
framework. Another strand of work considers robust policy learning
under ambiguity. \citet{kallus2021minimax} propose to learn an optimal
policy in the presence of partially identified treatment effects under
a sensitivity model. \citet{pu2021estimating} consider a minimax
regret policy for IV models under partial identification. \citet{cui2021individualized}
and \citet{d2021orthogonal} consider a variety of decision rules
in general settings where treatment effects are partially identified.
\citet{stoye2012minimax,yata2021optimal} develop finite-sample minimax
regret rules under partial identification of welfare. Moreover, \citet{han2023optimal}
proposes optimal dynamic treatment regimes through a partial welfare
ordering when the sequential randomization assumption is violated.
Policy learning under ambiguity is not limited to confounded settings.
There are other settings of robust decisions under ambiguity, for
example, when the treatment positivity assumption is violated \citep{ben2021safe},
when data sets are aggregated in meta analyses \citep{ishihara2021evidence}
and when the target population is shifted from the experiment population
\citep{adjaho2022externally}. The present paper contributes to this
literature of model ambiguity by considering a distributional welfare
that is partially identified.

There is also work focused on policy learning based on distributional
properties under point identification. \citet{leqi2021median} consider
the QTE as a criterion and \citet{qi2023robustness} consider maximizing
the average outcomes that are below a certain quantile. \citet{wang2018quantile,linn2017interactive}
consider maximizing the quantile of global welfare, which can be viewed
as a special case of \citet{kitagawa2021equality}. The latter study
considers estimating the optimal treatment allocation based on individual
characteristics when the objective is to maximize an equality-minded
rank-dependent welfare function, which essentially puts higher weights
on individuals with lower-ranked outcomes. Our work complements this
line of literature by introducing a different type of distributional
welfare using the distribution of treatment effects and proposing
decision-making under ambiguity. Further comparisons to this line
of work are made in Section \ref{sec:Treatment-Rules-and}. \citet{kock2022functional,kock2023treatment,kock2024regularizing} consider choosing an optimal treatment among a discrete set of treatments with distributional targets. Finally, \citet{manski2023statistical,kitagawa2023treatment} consider a distribution
or nonlinear function of regret and establish admissible treatment
rules within that framework. Although our welfare has distributional
aspects, when showing the theoretical guarantee of the estimated rules,
we use the standard the notion of the (mean) regret. 

\subsection{Organization of the Paper}

The paper is organized as follows. The next section formally introduces
our welfare criterion and compare it with criteria previously considered
in the literature. Then the minimax framework is proposed. Section
\ref{sec:Possible-Identifying-Assumptions} provides identifying assumptions
that can be used to narrow the bounds on the QoTE. Section \ref{sec:Theoretical-Properties-of}
presents the theoretical properties of the estimated policies for
constrained and unconstrained policy classes. Section \ref{sec:Calculating-Bounds}
discusses how to systematically calculate the bounds on the QoTE using
linear programming. Section \ref{sec:Empirical-Applications} presents
the two empirical applications. Finally, Section \ref{sec:Extensions}
concludes the paper by generalizing the paper's framework to other
related non-utilitarian welfare criteria. In the Supplemental Appendix,
Section A lists further identifying assumptions for tightening bounds
on the QoTE. Section B presents an additional empirical application,
and Section C contains numerical exercises. Section D further discusses
stochastic rules and Section E contains all proofs.

\section{Treatment Rules and Distributional Welfare\label{sec:Treatment-Rules-and}}

Let $Y\in\mathcal{Y}$ be the outcome, $X\in\mathcal{X}$ be covariates,
and $D\in\{0,1\}$ be binary treatment in respective supports. Let
$Y_{d}$ be the potential outcome that is consistent with the observed
outcome, that is, $Y=DY_{1}+(1-D)Y_{0}$. We define a treatment allocation
rule, or equivalently a \emph{policy}, as $\delta:\mathcal{X}\rightarrow\mathcal{A}\subseteq[0,1]$
where $\mathcal{A}$ is the action space. A deterministic rule corresponds
to $\mathcal{A}=\{0,1\}$ and a stochastic rule corresponds to $\mathcal{A}=[0,1]$.
Unless noted otherwise, we allow both in our general framework. Let
$\delta\in\mathcal{D}$ where $\mathcal{D}$ is the (potentially constrained)
space of $\delta$. For the allocation problem, a policymaker (PM)
would set an objective function that she maximizes to find the optimal
allocation rule.

\subsection{Introducing Distributional Welfare\label{subsec:Introducing-Distributional-Welfa}}

To motivate the objective function we propose, we first review the
most common objective function considered in the literature: the average
welfare.\footnote{Welfare is sometimes called a value function in the literature.}
The optimal policy under this welfare criterion can be defined as
$\delta_{ATE}^{*}\in\arg\max_{\delta\in\mathcal{D}}E[\delta(X)Y_{1}+(1-\delta(X))Y_{0}]$.
With deterministic rules in particular, the welfare can be written
as $E[\delta(X)Y_{1}+(1-\delta(X))Y_{0}]=E[Y_{\delta(X)}]$. See Section
E in the Appendix that shows how $E[\delta(X)Y_{1}+(1-\delta(X))Y_{0}]$
(and other welfare criteria appearing below) is compatible with stochastic
rules. Because $E[\delta(X)Y_{1}+(1-\delta(X))Y_{0}]=E[Y_{0}+\delta(X)(Y_{1}-Y_{0})]=E[Y_{0}]+E[\delta(X)E[Y_{1}-Y_{0}|X]],$
$\delta_{ATE}^{*}$ also satisfies 
\begin{align}
\delta_{ATE}^{*} & \in\arg\max_{\delta\in\mathcal{D}}E[\delta(X)E[Y_{1}-Y_{0}|X]],\label{eq:mean_welfare}
\end{align}
where the objective function corresponds to the \emph{welfare gain}.
Therefore, subject to the constraints, $\delta_{ATE}^{*}$ maximizes
the average of conditional average treatment effects (ATEs) either
chosen (in the case of deterministic policies) or weighted (in the
case of stochastic policies) by $\delta$, thus the notation ``$\delta_{ATE}^{*}$.''
For example, when $\mathcal{D}$ is not constrained, $\delta_{ATE}^{*}(x)=1\{E[Y_{1}-Y_{0}|X=x]\ge0\}$
for both deterministic and stochastic policies. In general, the formulation
\eqref{eq:mean_welfare} reveals an important fact: the conditional
treatment effect is the important basis for the policy choice. This
makes sense because the treatment should be allocated to those who
would benefit the most from it. This idea becomes important in introducing
our distributional welfare later.

Although it is the most common form of welfare, the average welfare
is obviously sensitive to outliers. For example, a small share of
individuals with $X=x$ and substantially large $Y_{1}-Y_{0}$ can
make $E[Y_{1}-Y_{0}|X=x]$ positive, suggesting to treat \emph{all}
individuals with $X=x$ even though the majority suffers from receiving
the treatment. This can be especially problematic when the distribution
of $Y_{1}-Y_{0}|X=x$ is skewed and heavy-tailed. This motivates us
to alternatively consider the quantile of individual treatment effects
$Y_{1}-Y_{0}$ (QoTE) as the basis for a welfare criterion (analogous
to \eqref{eq:mean_welfare}) and a corresponding optimal policy. Let
$Q_{\tau}(Y)\equiv\inf\{y:F_{Y}(y)\ge\tau\}$ be the $\tau$-quantile
of $Y$ and $Q_{\tau}(Y|X)\equiv\inf\{y:F_{Y|X}(y)\ge\tau\}$ be the
$\tau$-quantile of $Y$ conditional on $X$. We consider an optimal
policy that satisfies
\begin{align}
\delta^{*}\equiv\delta_{\tau}^{*} & \in\arg\max_{\delta\in\mathcal{D}}E[\delta(X)Q_{\tau}(Y_{1}-Y_{0}|X)],\label{eq:quantile_welfare}
\end{align}
where $Q_{\tau}(Y_{1}-Y_{0}|X)$ is the $\tau$-quantile of $Y_{1}-Y_{0}$
given $X$. That is, $\delta^{*}$ maximizes the average of conditional
QoTEs chosen (in the case of deterministic policies) or weighted (in
the case of stochastic policies) by $\delta$. With no constraint
in $\mathcal{D}$, $\delta^{*}(x)=1\{Q_{\tau}(Y_{1}-Y_{0}|X=x)\ge0\}$
for both deterministic and stochastic policies. The QoTE is less sensitive
to outliers than the ATE, so for example \eqref{eq:quantile_welfare}
with $\tau=0.5$ may be preferred to \eqref{eq:mean_welfare}. This
aspect makes the allocation decision within the $X=x$ group not driven
by treatment effects of a small share of individuals. In this sense,
this aspect of robustness can be viewed as the ``within-group robustness''
\citep{leqi2021median}. In general, $\tau$ (i.e., the rank in
individual treatment effects) represents individuals in that specific
quantile as a \emph{reference group} chosen by the PM. For example,
by choosing low $\tau$, the PM allocates the treatment only if most
individuals benefit from it because $Q_{\tau'}(Y_{1}-Y_{0}|X)\ge Q_{\tau}(Y_{1}-Y_{0}|X)$
for any $\tau'>\tau$. In other words, she ensures that disadvantaged
individuals with poor treatment effects are not harmed from receiving
the allocation. In this sense, low $\tau$ corresponds to a \emph{prudent
PM}. On the other hand, by choosing high $\tau$, the PM focuses on
benefiting solely the top-ranked individuals even though the majority
would suffer from it. In this sense, high $\tau$ corresponds to a
\emph{negligent PM}. Therefore, the choice of $\tau$ reflects the
level of prudence of the policy that the PM commits to.

The proposed optimal policy has another interesting interpretation
that relates to the PM's incentive. Let $\delta_{\tau}^{\dagger}\equiv1\{Q_{\tau}(Y_{1}-Y_{0}|X)\ge0\}\in\arg\max_{\delta:\mathcal{X\rightarrow\mathcal{A}}}E[\delta(X)Q_{\tau}(Y_{1}-Y_{0}|X)]$
be the first-best rule for $\mathcal{A}$ being \emph{either} $[0,1]$
or $\{0,1\}$. As mentioned above, $\delta_{\tau}^{\dagger}$ is an
optimal rule when no restriction is imposed on the class of $\delta$.
Suppose individuals who benefit from the treatment would vote for
it. Also suppose $\tau=0.5$. Then $\delta_{0.5}^{\dagger}(X)=1\{Q_{0.5}(Y_{1}-Y_{0}|X)\ge0\}$
can be viewed as a policy that obeys \emph{majority vote}. To see
this, note the following is true for continuously distributed $Y_{d}$:
$Q_{0.5}(Y_{1}-Y_{0}|X)\ge0$ if and only if $P[Y_{1}\ge Y_{0}|X]\ge P[Y_{1}<Y_{0}|X]$.
Therefore, the distributional welfare criterion \eqref{eq:quantile_welfare}
is consistent with a PM who has political incentive and whose decision
is influenced by vote shares. This interpretation can be generalized
by considering $Q_{0.5-\alpha/2}(Y_{1}-Y_{0}|X)\ge0$ for $0\le\alpha\le1$,
which is equivalent to $P[Y_{1}\ge Y_{0}|X]\ge P[Y_{1}<Y_{0}|X]+\alpha$
where $\alpha$ can be viewed as the vote share margin.

Exploring this interpretation further, we can show that the first-best
policy for the median can be viewed as the one that maximizes the
share of positively affected individuals or the correct classification
rate over a class of deterministic policies:

\begin{theorem}\label{thm:interpretation}Suppose $Y_{d}$ is continuously
distributed and $\mathcal{A}$ is either $[0,1]$ or $\{0,1\}$. Then,
the first best rule $\delta_{\tau}^{\dagger}(x)\equiv1\{Q_{\tau}(x)\ge0\}$
for $\tau=0.5$ satisfies
\begin{align}
\delta_{0.5}^{\dagger}\in\arg\max_{\delta:\mathcal{X}\rightarrow\mathcal{A}}E[\delta(X)Q_{0.5}(X)] & =\arg\max_{\delta:\mathcal{X}\rightarrow\{0,1\}}P\left[Y_{\delta(X)}-Y_{1-\delta(X)}>0\right]\label{eq:interpretation1}\\
 & =\arg\max_{\delta:\mathcal{X}\rightarrow\{0,1\}}P\left[\delta(X)\in\arg\max_{d}Y_{d}\right].\label{eq:interpretation2}
\end{align}
\end{theorem}

In the theorem, \eqref{eq:interpretation1} holds by the equivalence
result in the previous paragraph and \eqref{eq:interpretation2} is
immediate. Note that $P\left[\delta(X)\in\arg\max_{d}Y_{d}\right]$
is the correct classification rate. We can equivalently say that $\delta_{0.5}^{\dagger}$
minimizes the \emph{fraction negatively affected} by switching from
$1-\delta$ to $\delta$, namely, $P[Y_{\delta(X)}-Y_{1-\delta(X)}<0]$,
or the misclassification rate, $P\left[\delta(X)\notin\arg\max_{d}Y_{d}\right]$.
The latter extends \citet{kallus2022s}'s definition which focuses
on binary $Y_{d}$.

Related to the proposed welfare criterion, one can consider alternative
criteria that are robust to outliers. Focusing on a deterministic
policy (i.e., $\mathcal{A}=\{0,1\}$), \citet{wang2018quantile} consider
the marginal quantile of $Y_{\delta(X)}$ as their criterion, while
\citet{leqi2021median} focus on the average of conditional quantile
$Y_{\delta(X)}$. First, \citet{wang2018quantile} explore the optimal
policy under $Q_{\tau}(Y_{\delta(X)})$, which can be viewed as a
sensible quantity robust to outliers. Note that the randomness in
$Y_{\delta(X)}$ arises from both $Y_{d}$ and $X$. Because of that,
the optimal policy under $Q_{\tau}(Y_{\delta(X)})$ does not have
a closed form solution, which make the interpretation of the optimal
policy somewhat elusive. Moreover, \citet{leqi2021median} demonstrate
that the policy under this welfare criterion lacks ``across-group
fairness,'' in that the allocation decision for one group (defined
by $X=x$) can be influenced by the treatment effects of other groups
(defined by other $X=x'$). This issue stems from the difficulty in
associating the objective function $Q_{\tau}(Y_{\delta(X)})$ with
a clear notion of treatment effects or gains, unlike the other criteria
discussed in this section. To overcome this issue, \citet{leqi2021median}
consider the optimal policy under $E[Q_{\tau}(Y_{\delta(X)}|X)]$,
which achieves across-group fairness as $X$ is fixed in the calculation
of quantile. They show the optimal policy also satisfies $\delta_{QTE}^{*}\in E[\delta(X)\{Q_{\tau}(Y_{1}|X)-Q_{\tau}(Y_{0}|X)\}].$
That is, $\delta_{QTE}^{*}$ maximizes the average of conditional
QTEs chosen by $\delta$. However, a PM may find the allocation decision
based on the QTE undesirable because the individual at the $\tau$-quantile
of $Y_{1}$ may not be the same individual as the one at the $\tau$-quantile
of $Y_{0}$. Since introduced in \citet{doksum1974empirical} and
\citet{lehmann1975statistical}, the QTE has been a popular causal
parameter. However, its limitation is also acknowledged in the literature,
which seems more pronounced in the context of treatment allocation.
This aspect implies that it is difficult for the PM to aim the level
of prudence (e.g., to be conservative) as there is no clear notion
of a negligence or prudence associated with the level of $\tau$;
see Figure \ref{fig:application1-3} in the application (Section \ref{sec:Empirical-Applications})
for related discussions.

\subsection{Policies Robust to Model Ambiguity\label{subsec:Policies-Robust-to}}

Despite the desirable properties of our proposed objective function,
the main challenge of using \eqref{eq:quantile_welfare} as the welfare
criterion is that the QoTE is generally not point-identified even
under unconfoundedness. This is because, in general, the QoTE is not
equal to the QTE and, while the latter can be identified from
the marginal distributions of $Y_{1}$ and $Y_{0}$, the former can
only be identified from the joint distribution of $(Y_{1},Y_{0})$.
Therefore, we propose optimal policies that are robust to this ambiguity.
One may consider maximizing the worst-case gain: $\delta_{mmw}^{*}\in\arg\max_{\delta\in\mathcal{D}}\min_{F_{Y_{1},Y_{0}|X}\in\mathcal{F}}E[\delta(X)Q_{\tau}(Y_{1}-Y_{0}|X)],$
where $F_{Y_{1},Y_{0}|X}$ is the joint distribution of $(Y_{1},Y_{0})$
conditional on $X$ and $\mathcal{F}\equiv\mathcal{F}(P)$ is the
identified set of $F_{Y_{1},Y_{0}|X}$ given the data $P$. However,
this criterion is known to be overly pessimistic \citep{savage1951theory}.
Therefore, one may instead consider minimizing the worst-case regret:
\begin{align}
\delta_{mmr}^{*} & \in\arg\min_{\delta\in\mathcal{D}}\max_{F_{Y_{1},Y_{0}|X}\in\mathcal{F}}E[\{\delta^{\dagger}(X)-\delta(X)\}Q_{\tau}(Y_{1}-Y_{0}|X)],\label{eq:mmr}
\end{align}
where $\delta^{\dagger}\equiv\delta_{\tau}^{\dagger}\equiv1\{Q_{\tau}(Y_{1}-Y_{0}|X)\ge0\}$
is the first-best rule. The minimax regret criterion is free from
priors and thus avoids the feature of maximin mentioned above. The
two criteria becomes identical under point identification (i.e., when
$\mathcal{F}(P)$ is a singleton). Therefore, our primary focus is
the minimax policy.

For each $x$, define the identified interval for $Q_{\tau}(Y_{1}-Y_{0}|X=x)$
as 
\begin{align*}
[Q_{\tau}^{L}(x),Q_{\tau}^{U}(x)] & =\{Q_{\tau}(Y_{1}-Y_{0}|X=x):F_{Y_{1},Y_{0}|X}\in\mathcal{F}\}.
\end{align*}
Using these lower and upper bounds, we can derive closed-form expressions
for the inner optimization in \eqref{eq:mmr} (and similarly in the
objective function for $\delta_{mmw}^{*}$). To this end, we impose
a very weak assumption on the identified interval.

\begin{myas}{RC}\label{as:RC}The identified set $\mathcal{Q}(P)$
of $Q_{\tau}(Y_{1}-Y_{0}|X=\cdot)$ is rectangular, that is, $\mathcal{Q}(P) =\{Q_{\tau}(Y_{1}-Y_{0}|X=\cdot):Q_{\tau}(Y_{1}-Y_{0}|X=x)\in[Q_{\tau}^{L}(x),Q_{\tau}^{U}(x)]\}$.
\end{myas}

This assumption holds for the identified sets we derive in this paper.
It will be violated if one imposes certain shape restrictions on $Q_{\tau}(Y_{1}-Y_{0}|X=\cdot)$
such as monotonicity. We do not consider shape restrictions in this
paper as allowing for unrestricted heterogeneity across $X$ is important
in the context of optimal allocations. Essentially, this assumption
allows us to interchange the maximum or minimum over $\mathcal{F}$
with the expectation over $X$ \citep{kasy2016partial,d2021orthogonal}.\footnote{To illustrate this, consider a simple case of binary $X\in\{0,1\}$
and let $Q_{\tau}(x)\equiv Q_{\tau}(Y_{1}-Y_{0}|X=x)$ and $p_{x}\equiv P[X=x]$.
Then Assumption \ref{as:RC} imposes that $\{(Q_{\tau}(0),Q_{\tau}(1)):Q_{\tau}(x)\in[Q_{\tau}^{L}(x),Q_{\tau}^{U}(x)],x\in\{0,1\}\}$
is rectangular, which implies that, for example, 
\begin{align*} \min_{F_{Y_{1},Y_{0}|X}}E[\delta(X)Q_{\tau}(X)] & =p_{1}\delta(1)\min_{F_{Y_{1},Y_{0}|X}}Q_{\tau}(1)+p_{0}\delta(0)\min_{F_{Y_{1},Y_{0}|X}}Q_{\tau}(0)=E[\delta(X)\min_{F_{Y_{1},Y_{0}|X}}Q_{\tau}(X)].
\end{align*}
} Under Assumption \ref{as:RC}, we can easily show that $\delta_{mmr}^{*}$
equivalently satisfies 
\begin{align}
\delta_{mmr}^{*} & \in\arg\max_{\delta\in\mathcal{D}}E[\delta(X)\bar{Q}_{\tau}(X)],\label{eq:mmr2}
\end{align}
where $\bar{Q}_{\tau}(x)=Q_{\tau}^{U}(x)1\{Q_{\tau}^{L}(x)\ge0\}+Q_{\tau}^{L}(x)1\{Q_{\tau}^{U}(x)\le0\}+\left(Q_{\tau}^{U}(x)+Q_{\tau}^{L}(x)\right)1\{Q_{\tau}^{L}(x)<0<Q_{\tau}^{U}(x)\}.$
Also, we can show $\delta_{mmw}^{*}\in\arg\max_{\delta\in\mathcal{D}}E[\delta(X)Q_{\tau}^{L}(X)].$
In general, finding the optimal $\delta$ for \eqref{eq:mmr2} does
not yield a closed-form expression when the policy class $\mathcal{D}$
is constrained. Additionally, solving $\max_{\delta\in\mathcal{D}}E[\delta(X)\bar{Q}_{\tau}(X)]$
proves to be a challenging task as $\bar{Q}_{\tau}(\cdot)$ incorporates
an indicator function. Nonetheless, allowing the policy class to be
constrained is important because the PM may prefer a more parsimonious
rule (e.g., a linear rule) or be limited by certain institutional
constraints. Following \citet{zhao2012estimating}, we consider a
convex and continuous relaxation of \eqref{eq:mmr2} by utilizing
the hinge loss function $\phi(t)=\max(1-t,0)$ and introducing a regularization
term. This is done in Section \ref{subsec:Regret-Bounds-with-1} below.
The consistency of hinge loss is shown even when the class of $\delta$
is restricted \citep{kitagawa2021constrained}.

\section{Possible Identifying Assumptions\label{sec:Possible-Identifying-Assumptions}}

We now provide identifying assumptions that researchers may want to
consider imposing to shrink $\mathcal{F}$ (i.e., the identified set
for the joint distribution of $(Y_{1},Y_{0})$ conditional on $X$)
and thus $[Q_{\tau}^{L}(x),Q_{\tau}^{U}(x)]$. First, there are ways
to identify the marginal distribution of $Y_{d}$. The most obvious
approach is to impose conditional independence.

\begin{myas}{CI}[Conditional Independence]\label{as:CI}For $d\in\{0,1\}$,
$Y_{d}\perp D|X$.\end{myas}

Alternative to Assumption \ref{as:CI}, local copula modeling \citep{chernozhukov2024estimating}
or panel quantile regression models \citep{chernozhukov2013average}
can be used to identify $Q_{\tau}(Y_{d}|X)$. Given the identification
of the marginal distribution of $Y_{d}$, the \emph{Makarov bounds}
\citep{fan2010sharp} can be derived (see Section A), although they
tend to be uninformative. We now consider identifying assumptions
that can be used to yield tighter bounds, leading to more informative
decisions.

\begin{myas}{PD}[Positive Dependence]\label{as:PD}For $x\in\mathcal{X}$,
either (i) $P[Y_{1}\le y_{1},Y_{0}\le y_{0}|X=x]\le P[Y_{1}\le y_{1}|X=x]P[Y_{0}\le y_{0}|X=x]$,
(ii) $P[Y_{1}>y_{1}|Y_{0}>\cdot,X=x]$ and $P[Y_{0}>y_{0}|Y_{1}>\cdot,X=x]$
are non-decreasing and $P[Y_{1}\le y_{1}|Y_{0}\le\cdot,X=x]$ and
$P[Y_{0}\le y_{0}|Y_{1}\le\cdot,X=x]$ are non-increasing, or (iii)
$P[Y_{1}>y_{1}|Y_{0}=\cdot,X=x]$ and $P[Y_{0}>y_{0}|Y_{1}=\cdot,X=x]$
are non-decreasing, for all $y_{1},y_{0}\in\mathcal{Y}$.\end{myas}

Assumption \ref{as:PD} imposes various versions of positive dependence
between $Y_{1}$ and $Y_{0}$. This assumption makes sense when individuals
with high $Y_{1}$ (e.g., potential health with the treatment) tend
to have high $Y_{0}$ (e.g., potential health without the treatment)
and vice versa. \ref{as:PD}(iii), called \emph{stochastic increasingness} (SI), implies (ii), and (ii)
implies (i) \citep{joe2014dependence}. The following lemma provides a model for $Y_d$ that implies \ref{as:PD}: 
\begin{lemma}
    Assumption \ref{as:PD} holds if $Y_{d}=g(d,X,U)$
where $g(d,x,\cdot)$ is non-decreasing.
\end{lemma} In the previous example, $U$ may capture
underlying health conditions. The model assumption in the lemma trivially holds
with additively separable $U$ common in regression specification,
although it is substantially weaker than that. Due to its plausibility,
we consider this assumption as our leading one in later analyses. Maintaining Assumption \ref{as:CI},
Assumption \ref{as:PD} is helpful to obtain more informative bounds
on the conditional QoTE. For example, \citet{frandsen2021partial}
derive bounds on the distribution of treatment effects under an unconditional
version of \ref{as:PD}(iii). Instead of assuming positive dependence
between $Y_{1}$ and $Y_{0}$, one may want to impose stochastic dominance
of $Y_{d}$ between treatment and control groups or stochastic dominance
between $Y_{1}$ and $Y_{0}$ for each subgroup:

\begin{myas}{SD}[Stochastic Dominance]\label{as:SD}For $x\in\mathcal{X}$,
either (i) $P[Y_{d}\le y|D=1,X=x]\le P[Y_{d}\le y|D=0,X=x]$, or (ii)
$P[Y_{1}\le y|D=d,X=x]\le P[Y_{0}\le y|D=d,X=x]$.\end{myas}

Either under Assumption \ref{as:CI} or the existence of instrumental
variables (IVs), Assumption \ref{as:SD}(i) or \ref{as:SD}(ii) can
be used to narrow the bounds on the distribution of treatment effects
\citep{blundell2007changes,lee2021partial} and thus on
the QoTE.

Next, we present assumptions that help point-identify the conditional
QoTE. The following assumption is a special instance of Assumption
\ref{as:PD}.

\begin{myas}{RI}[Rank Invariance]\label{as:RI}For $d\in\{0,1\}$,
$Y_{d}=m_{d}(X,U_{d})$ where $m_{d}(x,\cdot)$ is strictly increasing
and $U_{d}|X=x$ is absolutely continuous and satisfies $U_{1}|_{X=x}=U_{0}|_{X=x}$
for given $x\in\mathcal{X}$. \end{myas}

\citet{heckman1997making} and \citet{chernozhukov2005iv} show the
identifying power of Assumption \ref{as:RI}. This assumption essentially
restricts heterogeneity by holding the ranks in $Y_{1}$ and $Y_{0}$
the same. This implies that, under this assumption, the QTE can be
interpreted as the difference between $Y_{1}$ and $Y_{0}$ for the
same individual. Yet, the QTE is \emph{not} identical to the OoTE
even under this assumption. Moreover, Assumption \ref{as:RI} implies
Assumption \ref{as:PD} because, suppressing $X$, $P[Y_{1}\le y_{1}|Y_{0}=y_{0}]=P[m_{1}(U)\le y_{1}|m_{0}(U)=y_{0}]=P[U\le m_{1}^{-1}(y_{1})|U=m_{0}^{-1}(y_{0})]$
and thus the probability is 1 when $y_{0}\le m_{0}(m_{1}^{-1}(y_{1}))$
and 0 otherwise. Under Assumptions \ref{as:CI} and \ref{as:RI},
$F_{\Delta|X}(\delta)$ is point identified. More generally, \citet{heckman1997making}
consider Markov kernels $M$ and $\tilde{M}$ so that $F_{Y_{1}|X}(y_{1})=\int M(y_{1},y_{0}|X)dF_{Y_{0}|X}(y_{0})$
and $F_{Y_{0}|X}(y_{0})=\int\tilde{M}(y_{1},y_{0}|X)dF_{Y_{1}|X}(y_{1})$.
Also see \citet{vuong2017counterfactual} for the case of endogenous
treatment with IVs. \citet[Section 2.5.3]{abbring2007econometric}
also consider perfect \emph{negative} dependence.\footnote{This corresponds to $U_{1}|_{X=x}=-U_{0}|_{X=x}$.}
Section A of the Supplemental Appendix contains an extended list of
assumptions for point identification, which includes deconvolution,
symmetry, and Roy models.

\section{Theoretical Properties of Estimated Policy\label{sec:Theoretical-Properties-of}}

Henceforth, let $Q_{\tau}(X)\equiv Q_{\tau}(Y_{1}-Y_{0}|X)$ for notational
simplicity. Focusing on the optimal policy $\delta_{mmr}^{*}$ based
on the minimax criterion, we provide theoretical guarantees for the
estimated policy. The policy can be readily estimated once the bounds
$[Q_{\tau}^{L}(X),Q_{\tau}^{U}(X)]$ on $Q_{\tau}(X)$ are estimated
using parametric or nonparametric methods with the sample of $(Y,D,X)$.
The theory includes the case of point identification as a special
case in which $Q_{\tau}(X)=Q_{\tau}^{L}(X)=Q_{\tau}^{U}(X)$.

Recall that our objective function is $V(\delta)\equiv E[\delta(X)Q_{\tau}(X)].$
To define the regret, we introduce a r.v. $A(x)$ that is distributed
as $Bernoulli(\delta(x))$. For a stochastic policy $\delta(x)\in[0,1]$,
$\delta(x)$ is the probability that $A(x)=1$. For a deterministic
policy $\delta(x)\in\{0,1\}$, the distribution of $A(x)$ is degenerate
and thus $A(x)=\delta(x)$. Define the regret of the ``classification''
as $R(\delta)\equiv V(\delta^{\dagger})-V(\delta)=E[|Q_{\tau}(X)|1\{A(X)\neq sign(Q_{\tau}(X))\}],$
where $\delta^{\dagger}(X)=1\{Q_{\tau}(X)\ge0\}$ and $sign(q)=1$
when $q\geq0$ and $sign(q)=0$ when $q<0$. Note that $R(\delta)$
is generally not point-identified and thus we define maximum regret
as $\bar{R}(\delta) \equiv\max_{Q_{\tau}(\cdot)\in[Q_{\tau}^{L}(\cdot),Q_{\tau}^{U}(\cdot)]}E[|Q_{\tau}(X)|1\{A(X)\neq sign(Q_{\tau}(X))\}]$. The maximum regret can be expressed in different ways, which are useful
in the analysis below.

\begin{lemma}\label{lem:max_regret}Suppose Assumption \ref{as:RC}
hold. For a stochastic or deterministic rule $\delta$, the maximum
regret can be expressed as
\begin{align}
\bar{R}(\delta) & =E\left[\max\left\{ [1-\delta(X)]\max(Q_{\tau}^{U}(X),0),\delta(X)\max(-Q_{\tau}^{L}(X),0)\right\} \right]\label{eq:max_regret1}\\
 & =-E[\delta(X)\bar{Q}_{\tau}(X)]+E\left[Q_{\tau}^{U}(X)1\{Q_{\tau}^{U}(X)\ge0\}\right]\label{eq:max_regret2}\\
 & =E[|\bar{Q}_{\tau}(X)|1\{A(X)\neq sign(\bar{Q}_{\tau}(X))\}]\nonumber \\
 & \qquad+E\bigg[\min(Q_{\tau}^{U}(X),-Q_{\tau}^{L}(X))1\{Q_{\tau}^{L}(X)<0<Q_{\tau}^{U}(X)\}\bigg].\label{eq:max_regret3}
\end{align}
\end{lemma}Note
that \eqref{eq:max_regret2} is used in expressing \eqref{eq:mmr2}.
Below, \eqref{eq:max_regret1} is used in Section \ref{subsec:Regret-Bounds-with}
and \eqref{eq:max_regret3} in Section \ref{subsec:Regret-Bounds-with-1}.
Now, we provide asymptotic bounds on these regrets evaluated at the
estimated stochastic and deterministic policies when $\mathcal{D}$
is unconstrained and constrained.

\subsection{Regret Bounds with Unconstrained Policy Class\label{subsec:Regret-Bounds-with}}

We assume that we are equipped with the consistent estimators for
$Q_{\tau}^{L}(X)$ and $Q_{\tau}^{U}(X)$.

\begin{myas}{EST}\label{as:EST}$Q_{\tau}(X)$ is bounded almost
surely, and $\hat{Q}_{\tau}^{L}(X)-Q_{\tau}^{L}(X)=o_{p}(1)$ and
$\hat{Q}_{\tau}^{U}(X)-Q_{\tau}^{U}(X)=o_{p}(1)$.\end{myas}

When $Q_{\tau}^{L}(X)$ and $Q_{\tau}^{U}(X)$ are known functions
of $F_{Y_{1}|X}$ and $F_{Y_{0}|X}$, Assumption \ref{as:EST} is
implied from the consistency of $\hat{F}_{Y_{1}|X}$ and $\hat{F}_{Y_{0}|X}$
by the continuous mapping theorem; see Section \ref{sec:Calculating-Bounds}
for the case of bounds that are computationally derived. Let $\delta^{*,stoch}\equiv\delta_{mmr}^{*,stoch}$
and $\delta^{*,determ}\equiv\delta_{mmr}^{*,determ}$ are the optimal
policies that minimize $\bar{R}(\delta)$ when $\delta$ is stochastic
and deterministic policies, respectively. Given the expression \eqref{eq:max_regret1},
a simple calculation yields 
\begin{align*}
\delta^{*,stoch}(x) & =\begin{cases}
1 & \text{if }Q_{\tau}^{L}(x)\ge0,\\
0 & \text{if }Q_{\tau}^{U}(x)\le0,\\
\frac{Q_{\tau}^{U}(x)}{Q_{\tau}^{U}(x)-Q_{\tau}^{L}(x)} & \text{if }Q_{\tau}^{L}(x)<0<Q_{\tau}^{U}(x),
\end{cases}
\end{align*}
and 
\begin{align*}
\delta^{*,determ}(x) & =\begin{cases}
1 & \text{if }Q_{\tau}^{L}(x)\ge0, \text{ or } \{Q_{\tau}^{L}(x)<0<Q_{\tau}^{U}(x)\text{ and }|Q_{\tau}^{L}(x)|<|Q_{\tau}^{U}(x)|\},\\
0 & \text{if }Q_{\tau}^{U}(x)\le0, \text{ or } \{Q_{\tau}^{L}(x)<0<Q_{\tau}^{U}(x)\text{ and }|Q_{\tau}^{L}(x)|>|Q_{\tau}^{U}(x)|\}.
\end{cases}
\end{align*}
Let $\hat{\delta}^{stoch}$ and $\hat{\delta}^{determ}$ are the estimates
of $\delta^{*,stoch}$ and $\delta^{*,determ}$, respectively.

\begin{theorem}\label{thm:regret_bound} Suppose Assumptions \ref{as:RC}
and \ref{as:EST} hold and $|Y|\le M$ for some constant $M$. The
regret of $\hat{\delta}^{stoch}$ is bounded by 
\begin{align*}
R(\hat{\delta}^{stoch}) & \leq E\left[\frac{Q_{\tau}^{L}(X)Q_{\tau}^{U}(X)}{Q_{\tau}^{L}(X)-Q_{\tau}^{U}(X)}1\{Q_{\tau}^{L}(X)<0<Q_{\tau}^{U}(X)\}\right]+o_{p}(1),
\end{align*}
where the ratio is defined to be $0$ whenever its denominator is
$0$. The regret of $\hat{\delta}^{determ}$ is bounded by
\begin{align*}
R(\hat{\delta}^{determ}) & \leq E\left[\min(\max(Q_{\tau}^{U}(X),0),\max(-Q_{\tau}^{L}(X),0))\right]+o_{p}(1).
\end{align*}
\end{theorem}

The proof of this theorem and all other proofs are collected in the
appendix. The leading term in each asymptotic regret bound collapses
to zero when either (i) the bounds on $Q_{\tau}(X)$ exclude zero
almost surely or (ii) $Q_{\tau}(X)$ is point-identified. These are
the situations in which we can identify the sign of $Q_{\tau}(X)$.
Recalling $\delta^{\dagger}(X)=1\{Q_{\tau}(X)\ge0\}$, this is enough
to achieve consistency $R\rightarrow0$ as the second term in each
regret bound is the sampling error. In general, the leading term becomes
larger as the endpoints $[Q_{\tau}^{L}(X),Q_{\tau}^{U}(X)]$ are farther
away from zero, which is intuitive. Finally, the leading term with
the stochastic rule ($\frac{Q_{\tau}^{L}(X)Q_{\tau}^{U}(X)}{Q_{\tau}^{L}(X)-Q_{\tau}^{U}(X)}$)
is weakly smaller than that with the deterministic rule ($\min(\max(Q_{\tau}^{U}(X),0),\max(-Q_{\tau}^{L}(X),0))$),
suggesting that the stochastic rule is more preferred when there is
model ambiguity. This is consistent with the findings in the literature
\citep{manski2007minimax,stoye2007minimax,cui2021individualized}. 

An immediate corollary of Theorem \ref{thm:regret_bound} establishes
the bound for the regret averaged over the sample of estimated policies.
Let $\mathbb{E}_{n}$ denote the expectation over the sample of $(Y,D,X)$.

\begin{corollary}Suppose Assumptions \ref{as:RC} and \ref{as:EST}
hold. Then, 
\begin{align*}
\mathbb{E}_{n}\left[R(\hat{\delta}^{stoch})\right] & \leq E\left[\frac{Q_{\tau}^{L}(X)Q_{\tau}^{U}(X)}{Q_{\tau}^{L}(X)-Q_{\tau}^{U}(X)}1\{Q_{\tau}^{L}(X)<0<Q_{\tau}^{U}(X)\}\right]+o_{p}(1),
\end{align*}
where the ratio is defined to be $0$ whenever its denominator is
$0$, and 
\begin{align*}
\mathbb{E}_{n}\left[R(\hat{\delta}^{determ})\right] & \leq E\left[\min(\max(Q_{\tau}^{U}(X),0),\max(-Q_{\tau}^{L}(X),0))\right]+o_{p}(1).
\end{align*}
\end{corollary}

\subsection{Regret Bounds with Constrained Policy Class\label{subsec:Regret-Bounds-with-1}}

As mentioned, allowing for a constrained policy class is crucial for
practical and institutional reasons. Our proposed method readily extends
to a scenario in which the policy class $\mathcal{D}$ is constrained.
Define the estimator of $\bar{Q}_{\tau}(\cdot)$ as $\widehat{\bar{Q}}_{\tau}(X)\equiv\hat{Q}_{\tau}^{U}(X)1\{\hat{Q}_{\tau}^{U}(X)\ge0\}+\hat{Q}_{\tau}^{L}(X)1\{\hat{Q}_{\tau}^{L}(X)\le0\}$
by noting that $\bar{Q}_{\tau}(x)$ also satisfies $\bar{Q}_{\tau}(x)=Q_{\tau}^{U}(x)1\{Q_{\tau}^{U}(x)\ge0\}+Q_{\tau}^{L}(x)1\{Q_{\tau}^{L}(x)\le0\}$.
We assume that the consistent estimators $\hat{Q}_{\tau}^{L}(X)$
and $\hat{Q}_{\tau}^{U}(X)$ are consistent with the specified rate
of convergence.

\begin{myas}{EST2}\label{as:EST2}$Q_{\tau}(X)$ is bounded almost
surely and $\hat{Q}_{\tau}^{L}(X)-Q_{\tau}^{L}(X)=O_{p}(n^{-\alpha})$
and $\hat{Q}_{\tau}^{U}(X)-Q_{\tau}^{U}(X)=O_{p}(n^{-\alpha})$ for
some $\alpha>0$.\end{myas}To overcome the computational problem
of obtaining $\delta_{mmr}^{*}$, we adopt the outcome weighted learning
framework \citep{zhao2012estimating}. We are interested in finding
a decision function $f:\mathcal{X}\rightarrow\mathbb{R}$ such that
$\delta(x)=1\{f(x)\ge0\}$. Note that by \eqref{eq:max_regret3},
we have $\bar{R}(f)\equiv\bar{R}(1\{f(\cdot)\ge0\})=E[|\bar{Q}_{\tau}(X)|1\{sign(f(X))\neq sign(\bar{Q}_{\tau}(X))\}]+E\bigg[\min(Q_{\tau}^{U}(X),-Q_{\tau}^{L}(X))1\{Q_{\tau}^{L}(X)<0<Q_{\tau}^{U}(X)\}\bigg].$
Accordingly, we define the surrogate regret as $\bar{R}^{S}(f)\equiv E[|\bar{Q}_{\tau}(X)|\phi\{sign(\bar{Q}_{\tau}(X))f(X)\}]+E\bigg[\min(Q_{\tau}^{U}(X),-Q_{\tau}^{L}(X))1\{Q_{\tau}^{L}(X)<0<Q_{\tau}^{U}(X)\}\bigg].$
Motivated by this expression, let $\hat{f}$ be the ML estimator of
$f$ from the following problem: 
\begin{align}
\hat{f} & =\arg\min_{f\in\mathcal{H}}\left\{ \frac{1}{n}\sum_{i=1}^{n}\left|\widehat{\bar{Q}}_{\tau}(X_{i})\right|\phi\{sign(\widehat{\bar{Q}}_{\tau}(X_{i}))f(X_{i})\}+\lambda_{n}||f||^{2}\right\} ,\label{eq:surrogate}
\end{align}
where $\phi(t)=\max\{1-t,0\}$ is the hinge loss, $\lambda_{n}$ is
the regularization parameter, and $||\cdot||$ is the norm in a function
space. We focus on the reproducing kernel Hilbert space (RKHS) $\mathcal{H}_{k}$
associated with Gaussian radial basis function kernels $k(x,z)=\exp(-\sigma_{n}^{2}||x-z||^{2})$.
By Theorem 2.1 of \citet{steinwart2007fast}, the complexity of $\mathcal{H}_{k}$
in terms of the covering number satisfies $\sup_{P_{n}}\log N\{B_{\mathcal{H}_{k}},\epsilon,L_{2}(P_{n})\}\leq c_{n}\epsilon^{-p},$
where $P_{n}$ is the distribution of $(Y,D,X)$, $c_{n}=c_{p,\delta,d}\sigma_{n}^{(1-p/2)(1+\delta)d}$,
$B_{\mathcal{H}_{k}}$ is the closed unit ball of $\mathcal{H}_{k}$,
$p\in(0,2]$, $\delta>0$, and $c_{p,\delta,d}$ is a constant. Define
the approximation error function as $a(\lambda_{n})=\inf_{f\in\mathcal{H}_{k}}E[|\bar{Q}_{\tau}(X)|\phi\{sign(\bar{Q}_{\tau}(X))f(X)\}+\lambda_{n}||f||^{2}]-\inf_{f}E[|\bar{Q}_{\tau}(X)|\phi\{sign(\bar{Q}_{\tau}(X))f(X)\},$
where the second infimum is over the unrestricted space of $f$. Note
that $a(\lambda_{n})$ goes to zero if the RKHS is rich enough. The
following theorem establishes the asymptotic bound on $\bar{R}(f)$.
The asymptotic bound on the true regret can be similarly obtained.

\begin{theorem}\label{thm:regret_bound_ML} Suppose Assumptions \ref{as:RC}
and \ref{as:EST2} hold, and suppose that $\lambda_{n}=o(1)$ and
$\lambda_{n}n^{\min(2\alpha,1)}\rightarrow\infty$. Then, with probability
larger than $1-\exp(-2\eta)$, we have 
\begin{align*}
\bar{R}(\hat{f})\leq & \inf_{f}\bar{R}(f)+a(\lambda_{n})+O_{p}(n^{-\alpha}\lambda_{n}^{-1/2})+M_{p}c_{n}^{\frac{2}{p+2}}n^{-\frac{2}{p+2}}\left(\lambda_{n}^{-\frac{2}{p+2}}+\lambda_{n}^{-1/2}\right)+\frac{K\eta}{n\lambda_{n}}(1+2\lambda_{n}^{1/2}),
\end{align*}
where $M_{p}$ and $K$ are constants. \end{theorem}

The leading term satisfies $\inf_{f}\bar{R}(f)=\bar{R}(f^{*})=E\left[\min(\max(Q_{\tau}^{U}(X),0),\max(-Q_{\tau}^{L}(X),0))\right]$,
because $f$ is not restricted and $f^{*}(x)=1$ if $Q_{\tau}^{L}(x)\ge0$,
$f^{*}(x)=0$ if $Q_{\tau}^{U}(x)\le0$ and $f^{*}(x)=sign(Q_{\tau}^{L}(x)+Q_{\tau}^{U}(x))$
if $Q_{\tau}^{L}(x)<0<Q_{\tau}^{U}(x)$. Note that this term coincides
with the leading term derived in Theorem \ref{thm:regret_bound} for
the deterministic rule. The second term is the approximation error
due to using the RKHS. The third term is the estimation error in estimating
the bounds. The rest of the terms are statistical errors in estimating
the policy.

\section{Calculating Bounds\label{sec:Calculating-Bounds}}

When $Q_{\tau}(X)$ is partially identified, we need a practical way
of calculating its bounds $[Q_{\tau}^{L}(x),Q_{\tau}^{U}(x)]=\{Q_{\tau}(x):F_{Y_{1},Y_{0}|X}\in\mathcal{F}\}.$
Unlike the Makarov bounds, the closed-form expression of the bounds
is not always available especially under Assumption \ref{as:PD}.
Therefore, it is fruitful to have a systematic procedure of calculating
the bounds. To this end, let $C(u_{1},u_{2}|X)$ be the copula for
$(U_{1},U_{2})\equiv(F_{Y_{1}}(Y_{1}),F_{Y_{0}}(Y_{0}))$ conditional
on $X$. By Sklar's Theorem, $C(u_{1},u_{2}|X)=F_{Y_{1},Y_{0}|X}(Q_{u_{1}}(Y_{1}|X),Q_{u_{2}}(Y_{0}|X))$.
Then, it satisfies $P[Y_{1}-Y_{0}\le t|X]=\int1\{Q_{u_{1}}(Y_{1}|X)-Q_{u_{2}}(Y_{0}|X)\le t\}dC(u_{1},u_{2}|X).$
Therefore, we can calculate the lower and upper bounds on the distribution
of $\Delta|X$ (recalling $\Delta\equiv Y_{1}-Y_{0}$) by 
\begin{align}
F_{\Delta|X}^{L}(t) & =\inf_{C(\cdot,\cdot|X)\in\mathcal{C}}\int1\{Q_{u_{1}}(Y_{1}|X)-Q_{u_{2}}(Y_{0}|X)\le t\}dC(u,v|X),\label{eq:LB}
\end{align}
and similarly for $F_{\Delta|X}^{U}(t)$ by taking supremum over $\mathcal{C}$,
where $\mathcal{C}$ is the class of copulas $C(\cdot,\cdot|X=x)$
restricted by identifying assumptions. Note that \eqref{eq:LB} can
be viewed as the (constrained version of the) Monge-Kantorovich problem
of finding the optimal coupling of marginal distributions in the optimal
transport theory \citep{villani2009optimal}. Then, for $\tau$-quantile
$Q_{\tau}$ of $\Delta$, we can obtain its lower and upper bounds
as $Q_{\tau}^{L}(X)=F_{\Delta|X}^{U,-1}(\tau)$ and $Q_{\tau}^{U}(X)=F_{\Delta|X}^{L,-1}(\tau)$,
where the inverse denotes the generalized inverse. In practice, \eqref{eq:LB}
is an infinite dimensional program, and thus infeasible. To transform
them into a linear program, we propose to approximate $C(u,v|x)$
using the Bernstein copula $C_{B}(u,v|x)$ \citep{sancetta2004bernstein}.

\begin{definition}[Bernstein Copula]For $j\in\{1,2\}$, let $P_{v_{j}}^{m_{j}}(u_{j})\equiv\left(\begin{array}{c}
m_{j}\\
v_{j}
\end{array}\right)u_{j}^{v_{j}}(1-u_{j})^{m_{j}-v_{j}}$. Then, $C_{B}:[0,1]^{2}\rightarrow[0,1]$ is a conditional Bernstein
copula for any $m_{j}\ge1$ and $x\in\mathcal{X}$ if $C_{B}(u_{1},u_{2}|x)=\sum_{v_{1}=0}^{m_{1}}\sum_{v_{2}=0}^{m_{2}}\beta\left(\frac{v_{1}}{m_{1}},\frac{v_{2}}{m_{2}},x\right)P_{v_{1}}^{m_{1}}(u_{1})P_{v_{2}}^{m_{2}}(u_{2})$
satisfies the usual properties of the copula function.\end{definition}

Then we can compute a feasible version of \eqref{eq:LB} as 
\begin{align}
 & \min_{\beta\in\mathcal{B}}\sum_{v_{1}=0}^{m_{1}}\sum_{v_{2}=0}^{m_{2}}\beta\left(\frac{v_{1}}{m_{1}},\frac{v_{2}}{m_{2}},X\right)\int 1\{Q_{u_{1}}(Y_{1}|X)-Q_{u_{2}}(Y_{0}|X)\le t\}dP_{v_{1}}^{m_{1}}(u_{1})dP_{v_{2}}^{m_{2}}(u_{2}),\label{eq:LB1-1}
\end{align}
and similarly for the upper bound by taking maximum over $\mathcal{B}$,
where $\mathcal{B}$ is the restricted set of $\beta(\cdot)$ to impose
identifying assumptions and guarantee that $C_{B}$ is a proper copula.
We omitted the latter restrictions for succinctness; see Theorem 2
in \citet{sancetta2004bernstein} for details. For example, to impose
Assumption \ref{as:PD}(iii) it is necessary to ensure that $C_{B}(u_{1}|u_{2},x)=\partial C_{B}(u_{1},u_{2},x)/\partial u_{2}$
and $C_{B}(u_{2}|u_{1},x)=\partial C_{B}(u_{1},u_{2},x)/\partial u_{1}$
are non-increasing. Then, by the desirable property of Bernstein,
this corresponds to $\beta\left(\frac{v_{1}}{m_{1}},\frac{v_{2}}{m_{2}},X\right)$
being weakly increasing in $v_{1}$ and $v_{2}$. The use of Bernstein
approximation for the systematic calculation of bounds on treatment
effects also appears in \citet{han2023optimal} and \citet{han2020sharp}
in different contexts. As an alternative to the Bernstein approximation,
one can discretize the space of $(U_{1},U_{2})\in[0,1]^{2}$ \citep{blundell2007changes,frandsen2021partial}.
This approach can be viewed as a simple local approximation involving
a uniform kernel. Finally, in practice, the inputs $Q_{u_{1}}(Y_{1}|X)$
and $Q_{u_{2}}(Y_{0}|X)$ of the linear program can be estimated using
standard nonparametric or parametric methods. When they are estimated
consistently, we can show that Assumption \ref{as:EST} holds for
the estimated outputs, $\hat{Q}_{\tau}^{L}(X)$ and $\hat{Q}_{\tau}^{U}(X)$,
of the linear program:

\begin{lemma}\label{lem:consistency}Suppose that, for $d\in\{0,1\}$,
$F_{Y_{d}|X}(y|X)$ and $Q_{\tau}(Y_{d}|X)$ are absolutely continuous
in $y\in\mathcal{Y}$ and $\tau\in(0,1)$, respectively, and $\hat{Q}_{\tau}(Y_{d}|X)$
is a consistent estimator of $Q_{\tau}(Y_{d}|X)$ for any $\tau\in(0,1)$,
almost surely. Then, $|\hat{Q}_{\tau}^{L}(X)-Q_{\tau}^{L}(X)|=o_{p}(1)$
and $|\hat{Q}_{\tau}^{U}(X)-Q_{\tau}^{U}(X)|=o_{p}(1)$.\end{lemma}

\section{Numerical Illustrations}

We numerically show the performance of treatment allocations across
welfare criteria, especially when the QoTE is partially identified.
For succincness, we only present a subset of results here; the full
results are contained in the Supplemental Appendix. As data-generating
processes (DGPs), we conside normal and lognormal distributions for
$(Y_{1},Y_{0})$ and Bernoulli for $D$. In simulation, the bounds
$Q_{\tau}^{L}$ and $Q_{\tau}^{U}$ are calculated under either no
assumption (i.e., Makarov bounds) or SI. For the policies $\delta_{mmr}^{*}$,
$\delta_{QTE}^{*}$ and $\delta_{ATE}^{*}$, we estimate their sample
counterparts $\hat{\delta}^{*}$, $\hat{\delta}_{QTE}^{*}$ and $\hat{\delta}_{ATE}^{*}$
by estimating $Q_{\tau}^{U}$, $Q_{\tau}^{L}$, $Q_{\tau}(Y_{d})$,
and $E[Y_{d}]$ ($d=0,1$).

Table \ref{tab:classification_rate1_short} presents the simulated
correct classification rates of the estimated policies. Under DGP
1 (with a normal distribution), both intervals under stochastic
increasingness (SI) (i.e., Assumption \ref{as:PD}(iii)) and no assumption
exclude $0$ and lie relatively far from it; therefore, both $\hat{\delta}$
and $\hat{\delta}^{SI}$ perform well. Under DGP 2 (with a log-normal
distribution),\emph{ }the interval under SI excludes $0$ while the
interval under no assumption covers $0$; therefore, $\hat{\delta}^{SI}$
performs better than $\hat{\delta}$; under this log-normal setting
and SI, $Q_{\tau}(Y_{1}-Y_{0})<E(Y_{1})-E(Y_{0})$ may be violated,
which occurs in the current subgroup and thus $\hat{\delta}^{SI}$
performs better than $\hat{\delta}_{ATE}$.

\begin{table}[ht]
\begin{centering}
\begin{tabular}{|c|cccccc|}
\hline 
{\scriptsize{}\diagbox{Optimal Policy}{Estimated Policy}} & {\scriptsize{}$\hat{\delta}^{stoch,SI}$} & {\scriptsize{}$\hat{\delta}^{stoch}$} & {\scriptsize{}$\hat{\delta}^{determ,SI}$} & {\scriptsize{}$\hat{\delta}^{determ}$} & {\scriptsize{}$\hat{\delta}_{QTE}$} & {\scriptsize{}$\hat{\delta}_{ATE}$}\tabularnewline
\hline 
\hline 
 & \multicolumn{6}{c|}{{\scriptsize{}DGP 1 (Normal)}}\tabularnewline
\hline 
{\scriptsize{}$\delta^{*}$} & {\scriptsize{}$100\%$} & {\scriptsize{}$100\%$} & {\scriptsize{}$100\%$} & {\scriptsize{}$100\%$} & {\scriptsize{}$1.5\%$} & {\scriptsize{}$100\%$}\tabularnewline
{\scriptsize{}$\delta_{QTE}^{*}$} & {\scriptsize{}$0\%$} & {\scriptsize{}$0\%$} & {\scriptsize{}$0\%$} & {\scriptsize{}$0\%$} & {\scriptsize{}$98.5\%$} & {\scriptsize{}$0\%$}\tabularnewline
{\scriptsize{}$\delta_{ATE}^{*}$} & {\scriptsize{}$100\%$} & {\scriptsize{}$100\%$} & {\scriptsize{}$100\%$} & {\scriptsize{}$100\%$} & {\scriptsize{}$1.5\%$} & {\scriptsize{}$100\%$}\tabularnewline
\hline 
 & \multicolumn{6}{c|}{{\scriptsize{}DGP 2 (Log-Normal)}}\tabularnewline
\hline 
{\scriptsize{}$\delta^{*}$} & {\scriptsize{}$99.5\%$} & {\scriptsize{}$48\%$} & {\scriptsize{}$100\%$} & {\scriptsize{}$32.5\%$} & {\scriptsize{}$100\%$} & {\scriptsize{}$4.5\%$}\tabularnewline
{\scriptsize{}$\delta_{QTE}^{*}$} & {\scriptsize{}$99.5\%$} & {\scriptsize{}$48\%$} & {\scriptsize{}$100\%$} & {\scriptsize{}$32.5\%$} & {\scriptsize{}$100\%$} & {\scriptsize{}$4.5\%$}\tabularnewline
{\scriptsize{}$\delta_{ATE}^{*}$} & {\scriptsize{}$0\%$} & {\scriptsize{}$52\%$} & {\scriptsize{}$0\%$} & {\scriptsize{}$67.5\%$} & {\scriptsize{}$0\%$} & {\scriptsize{}$95.5\%$}\tabularnewline
\hline 
\end{tabular}
\par\end{centering}
\caption{Correct Classification Rate ($\tau=0.25$, $n=1000$)}
\label{tab:classification_rate1_short}
\end{table}

\section{Empirical Applications\label{sec:Empirical-Applications}}

We consider two empirical applications to illustrate our method: (i)
the allocation of right heart catheterization to patients and (ii)
the allocation of job training to workers. We only present (i) here;
(ii) is contained in the Supplemental Appendix.

We consider the right heart catheterization (RHC) dataset from the
Study to Understand Prognoses and Preferences for Outcomes and Risks
of Treatments (SUPPORT) \citep{hirano2001estimation}. The treatment
$D$ in question is the RHC ($1$ if received and $0$ otherwise),
a diagnostic procedure for critically ill patients. The outcome $Y$
is the number of days from admission to death within 30 days (t3d30),
whose value ranges from 2 to 30. In contrast to the belief of practitioners
that the RHC is beneficial, studies like \citet{connors1996effectiveness}
found that patient survival is lower with the RHC than without. Therefore,
a relevant policy question in this critical situation is to find patients
for whom allocating (or avoiding) the RHC is life-saving. In the dataset,
5735 patients are divided into a treatment group (2184 patients) and
a control group (3551 patients). We consider the following covariates
as $X$: age, sex, coma in primary disease 9-level category (cat1\_coma),
coma in secondary disease 6-level category, (cat2\_coma), do not resuscitate
(DNR) status on day 1 (i.e., DNR when heart stops) (dnr1), estimated
probability of surviving 2 months (surv2md1), and APACHE III score
ignoring coma (i.e., ICU mortality score) (aps1).

\begin{figure}[H]
\begin{centering}
\centering \includegraphics[scale=0.3]{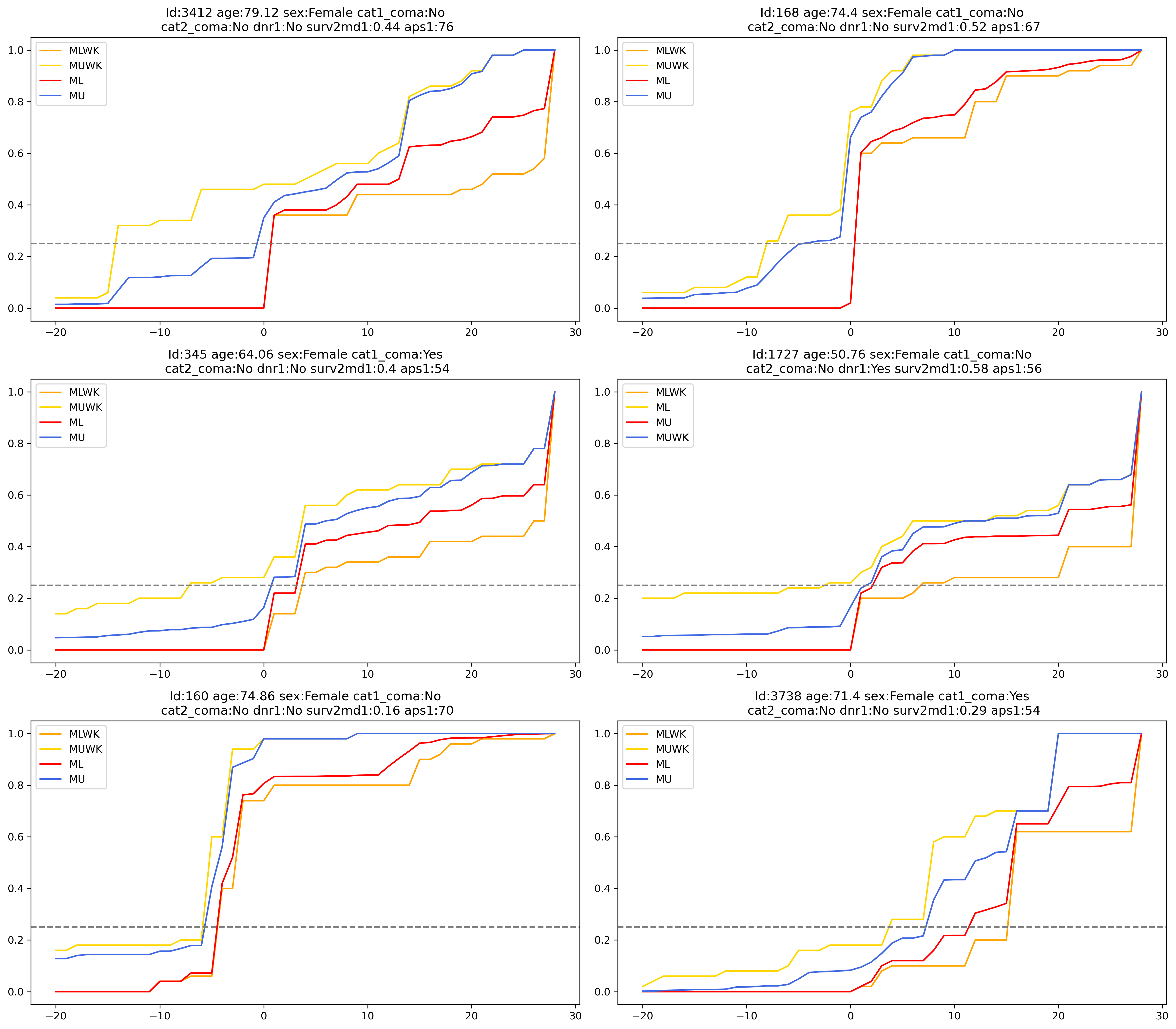}
\par\end{centering}
\begin{centering}
{\small \begin{tablenotes}
    \item \emph{Note}: In the figure, the vertical line indicates zero and the horizontal line indicates $\tau=0.25$.
\end{tablenotes}}
\par\end{centering}
\caption{Bounds on the QoTE of Six Representative Patients}
\label{fig:application1-1}
\end{figure}

To estimate the counterfactual distributions $F_{Y_{1}|X}$ and $F_{Y_{0}|X}$
of the outcome (t3d30) for different groups defined by the covariates,
we conduct a kernel regression in the treatment and control groups
separately with bandwidth under Scott's rule of thumb.\footnote{To simplify this process, we run the regression $P[Y<y_{j}|X=x]=E[1\{Y<y_{j}\}|X=x]$
on a series of $y_{j}=F_{Y}^{-1}(\frac{2j-1}{2k})$, where $k=1000$
and $j=1,...,k$.} Then we calculate the upper and lower bounds of the QoTE under SI (i.e., \ref{as:PD}(iii)) and no assumption
and make the decisions by using the proposed criterion based on the
QoTE. As shown in the simulation results in the Supplemental Appendix,
the SI and no-assumption bounds will not always give the same decisions,
and the information provided by the bounds differs from person to
person.

In Figure \ref{fig:application1-1}, we present six cases to show
the SI and no-assumption bounds of the QoTE. We only focus on deterministic
policies and $\tau=0.25$. These results illustrate how the actual
implementation of our proposed policies would look like for each individual.
It is shown that there is much heterogeneity in terms of the QoTEs
and thus the corresponding optimal decisions.

\begin{figure}[htbp]
\centering \subfigure[QoTE with $\tau=0.25$]{ \label{Fig.sub.1-1}
\includegraphics[width=3.9cm,height=3.9cm]{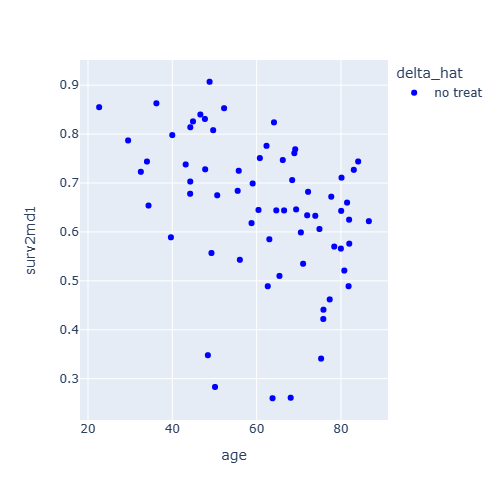}}
\subfigure[QoTE with $\tau=0.5$]{ \label{Fig.sub.2-1} \includegraphics[width=3.9cm,height=3.9cm]{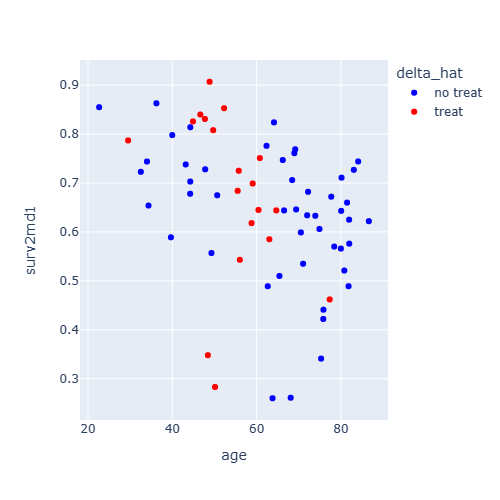}}
\subfigure[QoTE with $\tau=0.75$]{ \label{Fig.sub.3-1} \includegraphics[width=3.9cm,height=3.9cm]{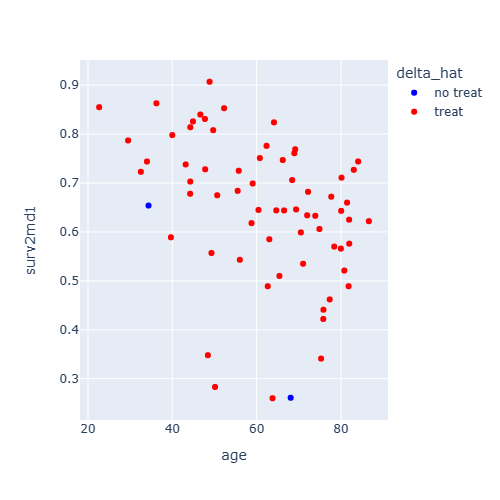}}
\caption{Treatment Decisions for Male Patients with Specific Health Conditions}
\label{fig:application1-2}
\end{figure}

Next, in Figure \ref{fig:application1-2}, we present the decisions
of allocating the RHC in terms of age and survival rate, which are
two important covariates for the allocation decision. We focus on
the male group whose primary and secondary disease categories are
not coma and APACHE score at day 1 is 54 and with resuscitate status.
We use the 0.25-quantile, median, and 0.75-quantile QoTE bounds to
represent prudent, majority-minded, and negligent PMs, respectively.
As expected, the 0.75-quantile bounds suggest the treatment option
more often than the bounds with the other quantile probabilities.
Given that the 0.25-quantile bounds suggest the most prudent decisions,
the suggested treatment option can be viewed as a compelling recommendation.
\begin{figure}[htbp]
\centering \subfigure[QTE with $\tau=0.25$]{ \label{Fig2.sub.1-1}
\includegraphics[width=3.9cm,height=3.9cm]{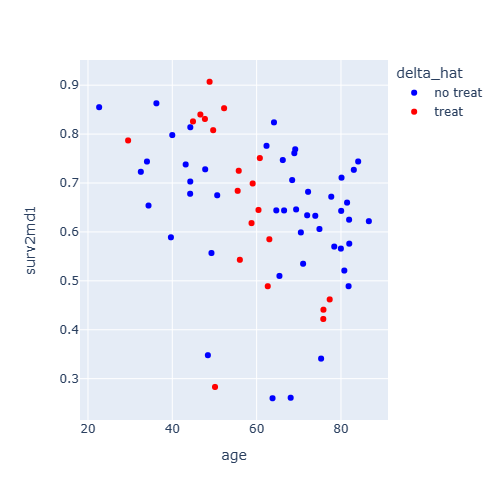}}
\subfigure[QTE with $\tau=0.5$]{ \label{Fig2.sub.2-1} \includegraphics[width=3.9cm,height=3.9cm]{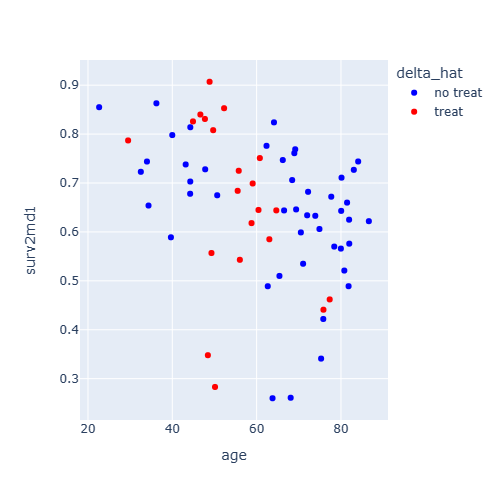}}
\subfigure[QTE with $\tau=0.75$]{ \label{Fig2.sub.3-1} \includegraphics[width=3.9cm,height=3.9cm]{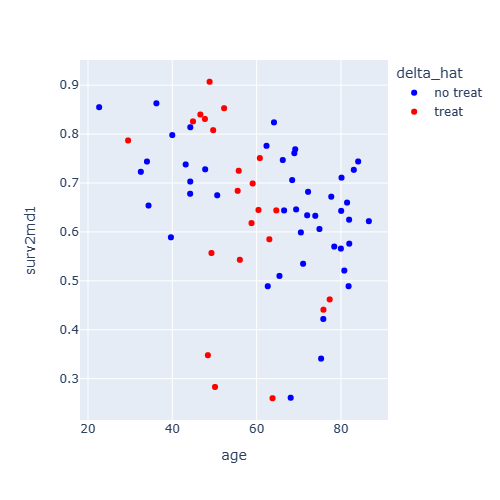}}
\subfigure[ATE]{ \label{Fig2.sub.4-1} \includegraphics[width=3.9cm,height=3.9cm]{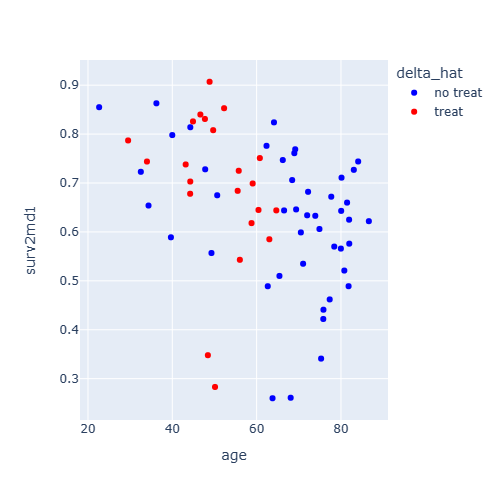}}
\caption{Treatment Decisions Based on the QTEs and the ATE}
\label{fig:application1-3}
\end{figure}

For comparison, in Figure \ref{fig:application1-3}, we present the
allocation decisions based on the 0.25-quantile, median, and 0.75-quantile
QTE and the ATE. Interestingly, there is no obvious tendency in decisions
when the quantile probability increases from 0.25 to 0.75, which reflects
the possible limitation of using the QTE as the basis for decisions
(e.g., the quantile probability does not capture the level of prudence).
The decisions based on the ATE show how they can be viewed as the
most common approach in the literature. They look very similar to
the decisions based on the median QoTE bounds, although there are
a few points that differ from the latter. Note that the policy based
on the median QoTE bounds can be viewed as a robustness check for
the policy based on the ATE.

\section{Generalization\label{sec:Extensions}}

The joint distribution of the potential outcomes may contain other
useful information about treatment effect heterogeneity for policy
learning. The theoretical results of this paper can be generalized
to any setting where welfare is defined as a functional of the joint
distribution of potential outcomes. Consider an optimal rule that
satisfies
\begin{align}
\delta^{**} & \in\arg\max_{\delta\in\mathcal{D}}E\left[\delta(X)\Lambda(F_{Y_{1},Y_{0}|X})\right],\label{eq:extension0}
\end{align}
where $\Lambda:\tilde{\mathcal{F}}\rightarrow\mathbb{R}$ is some
functional of the joint distribution of $(Y_{1},Y_{0})$ given $X$.
Our original criterion \eqref{eq:quantile_welfare} is a special case
of \eqref{eq:extension0} with $\Lambda(F_{Y_{1},Y_{0}|X})=Q_{\tau}(Y_{1}-Y_{0}|X)$.
We propose other examples of the criterion $\Lambda(F_{Y_{1},Y_{0}|X})$
that may interest a non-utilitarian PM.

\begin{example}\label{ex1}Consider
\begin{align}
\delta_{\bar{y}}^{**} & \in\arg\max_{\delta\in\mathcal{D}}E\left[\delta(X)E[Y_{1}-Y_{0}|Y_{0}<\bar{y},X]\right]\label{eq:extension1}
\end{align}
for some predetermined $\bar{y}$. This can be motivated by maximizing
the average of $\delta(X)Y_{1}+(1-\delta(X))Y_{0}$ (or simply $Y_{\delta(X)}$
with deterministic $\delta$), conditional on $Y_{0}<\bar{y}$: $E\left[\delta(X)Y_{1}+(1-\delta(X))Y_{0}|Y_{0}<\bar{y}\right]$,
because $E[\delta(X)Y_{1}+(1-\delta(X))Y_{0}|Y_{0}<\bar{y}]=E[Y_{0}|Y_{0}<\bar{y}]+E[\delta(X)E[Y_{1}-Y_{0}|Y_{0}<\bar{y},X]]$.
The PM with the latter criterion focuses on the welfare
of a disadvantaged population, defined by the baseline outcome $Y_{0}$
being less than $\bar{y}$. A similar intuition applies to the criterion
\eqref{eq:extension1}, which can be interpreted as addressing the
average gain for the disadvantaged. The ATE for the disadvantaged,
$E[Y_{1}-Y_{0}|Y_{0}<\bar{y}]$, appears as a parameter of interest
in the context of policy evaluation \citep{heckman1995assessing},
which is adapted here for the context of optimal allocation; also
see \citet{kaji2023assessing} for recent related work. \end{example}

\begin{example}\label{ex2}Alternative to Example \ref{ex1}, one
can consider $\Lambda(F_{Y_{1},Y_{0}|X})=Q_{\tau}(Y_{1}-Y_{0}|Y_{0}<\bar{y},X)$.
This would make the criterion robust to outliers and add an additional
dimension, $\tau$, to target a specific subgroup. Analogous to Theorem
\ref{thm:interpretation}, for continuously distributed $Y_{d}$ and
deterministic policy $\delta$, the first-best policy under $Q_{0.5}(Y_{1}-Y_{0}|Y_{0}<\bar{y},X)$
is the one that maximizes $P\left[Y_{\delta(X)}-Y_{1-\delta(X)}>0|Y_{0}<\bar{y}\right]=P\left[\delta\in\arg\max_{d}Y_{d}|Y_{0}<\bar{y}\right]$.\end{example}

\begin{example}\label{ex3}One may wish to target individuals worst-affected
by the treatment rather than those who are worst off at baseline.
The conditional value at risk (CVaR) of the distribution of individual
treatment effects can serve just that: $\Lambda(F_{Y_{1},Y_{0}|X})=E[Y_{1}-Y_{0}|Y_{1}-Y_{0}<\bar{\Delta},X]$.
\citet{kallus2023treatment} considers the CVaR, $E[Y_{1}-Y_{0}|Y_{1}-Y_{0}<\bar{\Delta}]$
with $\bar{\Delta}=Q_{\tau}(Y_{1}-Y_{0})$, as a parameter related
to distributional treatment effects and provides a sharp upper bound
and, under restricted heterogeneity, a sharp lower bound on the CVaR.\end{example}

In all these examples, $\Lambda(F_{Y_{1},Y_{0}|X})$ is not generally
point-identified, so one can follow the approach in Section \ref{subsec:Policies-Robust-to}
by considering $\delta_{mmr}^{**} \in\arg\min_{\delta\in\mathcal{D}}\max_{F_{Y_{1},Y_{0}|X}\in\mathcal{F}}E\left[\delta(X)\Lambda(F_{Y_{1},Y_{0}|X})\right]$. Then, one can apply the identifying assumptions listed in Section
\ref{sec:Possible-Identifying-Assumptions} and the computational
method proposed in Section \ref{sec:Calculating-Bounds} to systematically
calculate the bounds on $\Lambda(F_{Y_{1},Y_{0}|X})$ and to eventually
estimate $\delta_{mmr}^{**}$. Let $\Lambda^{L}(X)$ and $\Lambda^{U}(X)$
be the lower and upper bounds on $\Lambda(F_{Y_{1},Y_{0}|X})$. Then
the theoretical properties of the estimated $\delta_{mmr}^{**}$ with
constrained and unconstrained policy classes can be established based
on Section \ref{sec:Theoretical-Properties-of} by simply replacing
$Q_{\tau}^{L}(X)$ and $Q_{\tau}^{U}(X)$ with $\Lambda^{L}(X)$ and
$\Lambda^{U}(X)$ throughout the section.

\newpage
\begin{center}
    \LARGE Supplemental Appendix\\[1ex]
\end{center}

\appendix

In this appendix, Section \ref{sec:Possible-Identifying-Assumptions} lists further
identifying assumptions for tightening bounds on the QoTE. Section
\ref{sec:Additional-Empirical-Application} presents an additional
empirical application, and Section \ref{sec:Numerical-Illustrations}
contains numerical exercises. Section \ref{sec:Welfare-Criteria-with}
further discusses stochastic rules, and Section \ref{sec:Proofs}
contains proofs of the theorems and lemmas.

\section{More on Possible Identifying Assumptions\label{sec:Possible-Identifying-Assumptions}}

Recall the bounds on the QoTE:
\begin{align*}
[Q_{\tau}^{L}(x),Q_{\tau}^{U}(x)] & =\{Q_{\tau}(Y_{1}-Y_{0}|X=x):F_{Y_{1},Y_{0}|X}\in\mathcal{F}\}.
\end{align*}
Continuing Section 3 in the main text, we provide a menu of identifying
assumptions that researchers may want to consider imposing to shrink
$\mathcal{F}$ (i.e., the identified set for the joint distribution
of $(Y_{1},Y_{0})$ conditional on $X$). This would consequently
tighten $[Q_{\tau}^{L}(x),Q_{\tau}^{U}(x)]$, and sometimes reduce
it to a singleton, achieving point identification. First, there are
ways to identify the marginal distribution of $Y_{d}$. We restate
Assumption \ref{as:CI} here:

\begin{myas}{CI}[Conditional Independence]\label{as:CI}For $d\in\{0,1\}$,
$Y_{d}\perp D|X$.\end{myas}

An clear example where this assumption holds is when data from randomized
experiments are available. In general, one can argue that the treatment
is exogenous after adequately controlling for covariates. Alternative
to Assumption \ref{as:CI}, local copula modeling \citep{chernozhukov2024estimating}
or panel quantile regression models \citep{chernozhukov2013average}
can be used to identify $Q_{\tau}(Y_{d}|X)$.

The identification of the marginal distribution of $Y_{d}$ yields
bounds on the QoTE, $Q_{\tau}(Y_{1}-Y_{0}|X=x)$. The best-known sharp
bounds on the QoTE are derived by \citet{makarov1982estimates} and
\citet{williamson1990probabilistic} without imposing further restrictions
on the data-generating mechanism. Under Assumption \ref{as:CI}, the
\emph{Makarov bounds} can be used, which we derive here by trivially
extending Lemma 2.3 in \citet{fan2010sharp} to incorporate covariates.\footnote{The Makarov bounds are \emph{not} achieved at the Fr\'echet-Hoeffding
bounds for the joint distribution of $(Y_{1},Y_{0})$ \citep[Lemma 2.1]{fan2010sharp}.
Also, the bounds are point-wise but not uniformly sharp \citep{firpo2019partial}.}

\begin{proposition}For $0\le\tau\le1$, $Q_{\tau}^{L}(x)\le Q_{\tau}(Y_{1}-Y_{0}|X=x)\le Q_{\tau}^{U}(x)$
where 
\begin{align*}
Q_{\tau}^{L}(x) & =\begin{cases}
\sup_{u\in[0,\tau]}[Q_{u}(Y_{1}|X=x)-Q_{1+u-q}(Y_{0}|X=x)] & \text{if }q\neq1\\
Q_{1}(Y_{1}|X=x)-Q_{0}(Y_{0}|X=x) & \text{if }q=1.
\end{cases}\\
Q_{\tau}^{U}(x) & =\begin{cases}
\inf_{u\in[\tau,1]}[Q_{u}(Y_{1}|X=x)-Q_{u-q}(Y_{0}|X=x)] & \text{if }q\neq0\\
Q_{0}(Y_{1}|X=x)-Q_{1}(Y_{0}|X=x) & \text{if }q=0,
\end{cases}
\end{align*}
\end{proposition}

It is known that the Makarov bounds tend to be uninformative. We consider
a range of identifying assumptions that can be used to yield tighter
bounds, leading to more informative decisions. The following assumptions
appear in the main text:

\begin{myas}{PD}[Positive Dependence]\label{as:PD}For $x\in\mathcal{X}$,
either (i) $P[Y_{1}\le y_{1},Y_{0}\le y_{0}|X=x]\le P[Y_{1}\le y_{1}|X=x]P[Y_{0}\le y_{0}|X=x]$,
(ii) $P[Y_{1}>y_{1}|Y_{0}>\cdot,X=x]$ and $P[Y_{0}>y_{0}|Y_{1}>\cdot,X=x]$
are non-decreasing and $P[Y_{1}\le y_{1}|Y_{0}\le\cdot,X=x]$ and
$P[Y_{0}\le y_{0}|Y_{1}\le\cdot,X=x]$ are non-increasing, or (iii)
$P[Y_{1}>y_{1}|Y_{0}=\cdot,X=x]$ and $P[Y_{0}>y_{0}|Y_{1}=\cdot,X=x]$
are non-decreasing, for all $y_{1},y_{0}\in\mathcal{Y}$.\end{myas}

\begin{myas}{SD}[Stochastic Dominance]\label{as:SD}For $x\in\mathcal{X}$,
either (i) $P[Y_{d}\le y|D=1,X=x]\le P[Y_{d}\le y|D=0,X=x]$, or (ii)
$P[Y_{1}\le y|D=d,X=x]\le P[Y_{0}\le y|D=d,X=x]$.\end{myas}

Next, we present assumptions that help point-identify the conditional
QoTE. The first assumption appears in the main text:

\begin{myas}{RI}[Rank Invariance]\label{as:RI}For $d\in\{0,1\}$,
$Y_{d}=m_{d}(X,U_{d})$ where $m_{d}(x,\cdot)$ is strictly increasing
and $U_{d}|X=x$ is absolutely continuous and satisfies $U_{1}|_{X=x}=U_{0}|_{X=x}$
for given $x\in\mathcal{X}$. \end{myas}

\begin{myas}{SY}[Symmetric Distribution]\label{as:SY}The distribution
of $Y_{1}-Y_{0}|X$ is symmetric.\end{myas}

Under this assumption, $Q_{0.5}(Y_{1}-Y_{0}|X)=E[Y_{1}-Y_{0}|X]$,
which is point-identified under Assumption \ref{as:CI}.

\begin{myas}{DC}[Deconvolution]\label{as:DC}$Y_{1}-Y_{0}\perp Y_{0}|X$.\end{myas}

\citet{heckman1995assessing} show how Assumption \ref{as:DC} can
be useful to point identify $F_{Y_{1},Y_{0}|X}$ when combined with
the following assumption, which is stronger than Assumption \ref{as:CI}.\footnote{Interested readers can refer to Section 2.5.5 of \citet{abbring2007econometric}
to see how Assumption \ref{as:DC} relates to a normal random coefficient
model.}

\begin{myas}{CI2}[Joint Conditional Independence]\label{as:CI2}$(Y_{1},Y_{0})\perp D|X$.\end{myas}

The next set of assumptions explicitly posits that the treatment selection
is determined by the net gain from the treatment.

\begin{myas}{RY}[Roy Model]\label{as:RY}$D=1\{Y_{1}\ge Y_{0}\}$
and $X=(X_{0},X_{1},X_{c})$ where (i) $Y_{1}=g_{1}(X_{1},X_{c})+U_{1}$
and $Y_{0}=g_{0}(X_{0},X_{c})+U_{0}$, (ii) $(U_{0},U_{1})\perp(X_{0},X_{1},X_{c})$,
(iii) $(U_{0},U_{1})$ are absolutely continuous with $\text{Supp}(U_{0},U_{1})=\mathbb{R}^{2}$,
(iv) for each $X_{c}$ and $d\in\{0,1\}$, $g_{d}(X_{d},X_{c}):\mathbb{R}^{k_{d}}\rightarrow\mathbb{R}$
for all $X_{1-d}$, $\text{Supp}(g_{d}(X_{d},X_{c})|X_{c},X_{1-d})=\mathbb{R}$
for all $X_{c},X_{1-d}$, and $\text{Supp}(X_{d}|X_{1-d},X_{c})=\text{Supp}(X_{d})=\mathbb{R}$
for all $X_{c},X_{1-d}$, and (v) for $d\in\{0,1\}$, $U_{d}$ has
zero median.\end{myas}

Under Assumption \ref{as:RY}, $g_{0}$, $g_{1}$, and $F_{U_{0},U_{1}}$
are point identified (\citet[Theorem A-1]{heckman1995assessing});
see \citet{heckman1990empirical} for Gaussian case.

\begin{myas}{RY2}[Extended Roy Model]\label{as:RY2}$D=1\{Y_{1}\ge h(Y_{0},X,Z)\}$
where (i) $(Y_{0},Y_{1})\perp Z|X$, (ii) $\text{Supp}(Y_{0},Y_{1}|X)=\mathbb{R}^{2}$,
(iii) $h(y_{0},x,\cdot)$ and $h(\cdot,x,z)$ are strictly increasing
for any $(y_{0},x,z)$, and (iv) $h(y_{0},x,\cdot)$ is differentiable.\end{myas}

Under Assumption \ref{as:RY2}, \citet{lee2023nonparametric} show
that $F_{Y_{1},Y_{0}|X}(y_{1},y_{0}|x)$ is point identified for $(y_{1},y_{0})\in\mathcal{H}(x)$
where $\mathcal{H}(x)\equiv\{(y_{1},y_{0})\in\mathbb{R}^{2}:y_{1}=h(y_{0},x,z)\text{ for some }z\in\text{Supp}(Z|X=x)\}$.
Its implication for our purpose is that $Q_{\tau}(Y_{1}-Y_{0}|X=x)$
is identified if and only if $\{(y_{1},y_{0})\in\mathbb{R}^{2}:y_{1}-y_{0}=Q_{\tau}(Y_{1}-Y_{0}|X=x)\}\subseteq\mathcal{H}(x)$.

\section{Additional Empirical Application\label{sec:Additional-Empirical-Application}}

\subsection{Application II: Allocation of Job Training\label{subsec:Application-II:-Allocation}}

The dataset is collected from the National Job Training Partnership
Act (JTPA) Study \citep{bloom1997benefits}. We use a subset that
includes 9,223 adults; 6,133 of them received job training, while
the remaining 3,090 did not. The treatment $D$ in question is the
job training. In this experiment, we use the 30-month earnings after
the job training program as the measure of outcome $Y$ and the sex,
years of education, high school diploma, and previous earnings in
\$10K before the program as covariates $X$. Based on the data, the
kernel regression has been conducted in the treatment and control
groups separately to obtain the $\hat{F}_{Y_{1}|X}$ and $\hat{F}_{Y_{0}|X}$.
From the estimated conditional distributions, we obtain the upper
and lower bounds under SI (i.e., Assumption \ref{as:PD}(iii)) and
no assumption for each individual.
\begin{table}
\begin{centering}
\begin{tabular}{|c|c|c|c|c|}
\hline 
Worker ID & $(Q_{\tau}^{L},Q_{\tau}^{U})$ & $\hat{\delta}$ & $(Q_{\tau}^{L,SI},Q_{\tau}^{U,SI})$ & $\hat{\delta}^{SI}$\tabularnewline
\hline 
\hline 
317237 & $(-14.27,0.69)$ & $0$ & $(-1.44,0.83)$ & $0$\tabularnewline
\hline 
310493 & $(-8.07,0.40)$ & $0$ & $(-0.37,0.45)$ & $1$\tabularnewline
\hline 
302861 & $(-7.17,3.69)$ & $0$ & $(0.13,0.44)$ & $1$\tabularnewline
\hline 
302890 & $(-2.50,6.75)$ & $1$ & $(-2.23,-0.02)$ & $0$\tabularnewline
\hline 
305160 & $(-5.88,-4.44)$ & $0$ & $(-3.09,-2.48)$ & $0$\tabularnewline
\hline 
303890 & $(3.7,15.12)$ & $1$ & $(-3.21,-1.19)$ & $0$\tabularnewline
\hline 
\end{tabular}
\par\end{centering}
\caption{Bounds on the QoTE ($\tau=0.25$) and Estimated Policies}
\label{tab:application2}
\end{table}
\begin{figure}[H]
\begin{centering}
\centering \includegraphics[scale=0.3]{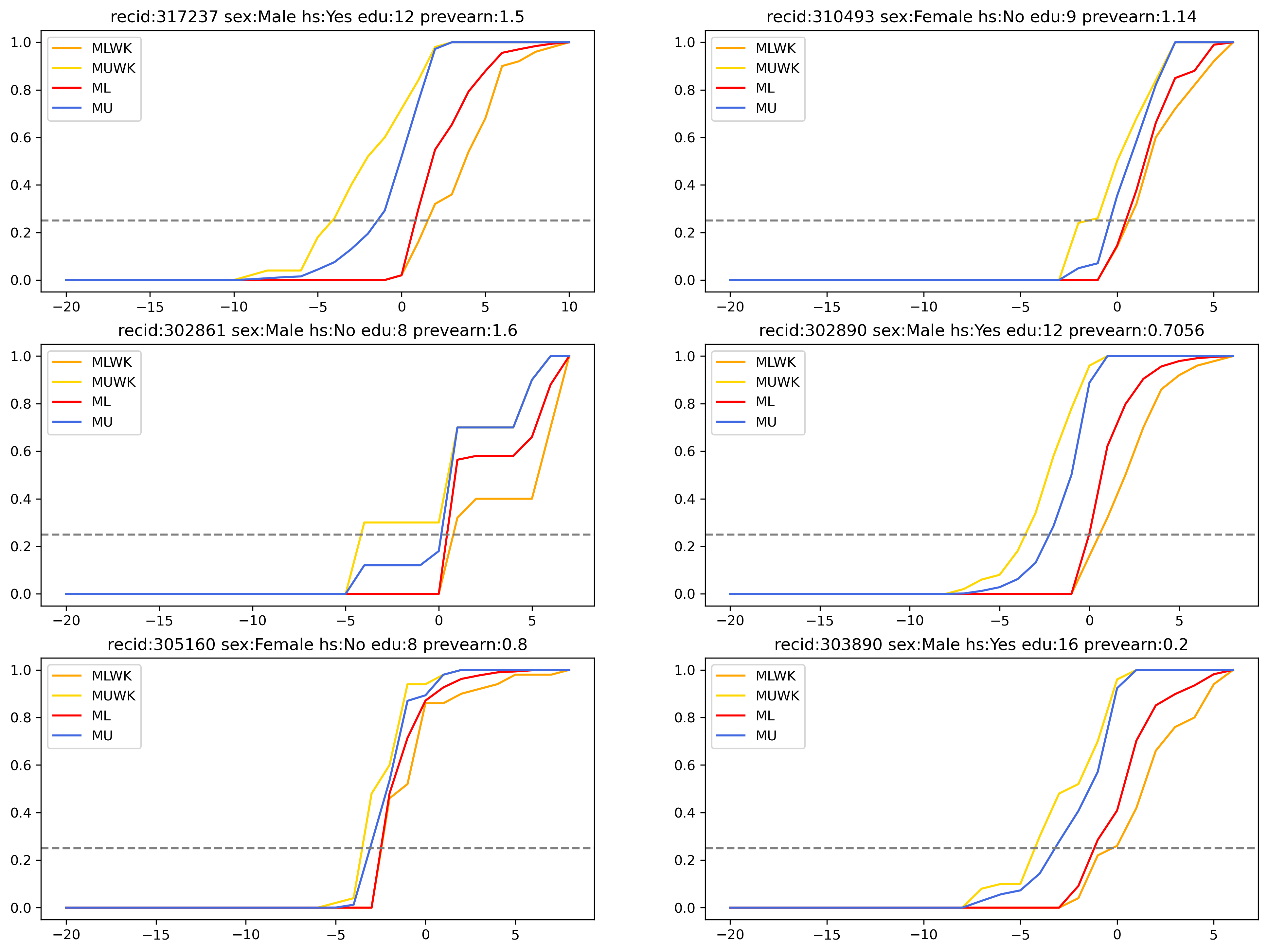}
\par\end{centering}
{\small \begin{tablenotes}
    \item \emph{Note}: In the figure, the vertical line indicates zero and the horizontal line indicates $\tau=0.25$.
\end{tablenotes}}\caption{Bounds on the QoTE of Six Representative Workers}
\label{fig:application2-1}
\end{figure}

In Table \ref{tab:application2} and Figure \ref{fig:application2-1},
we present six cases and their covariates to show bounds on the QoTE
(i.e., the effect of job training on earnings) under SI and no assumption.
Similar to the first application, we find heterogeneity in the treatment
effects and thus the optimal decisions, but less so than the first
application.
\begin{figure}[htbp]
\centering \subfigure[QoTE with $\tau=0.25$]{ \label{Fig.sub.1}
\includegraphics[width=3.9cm,height=3.9cm]{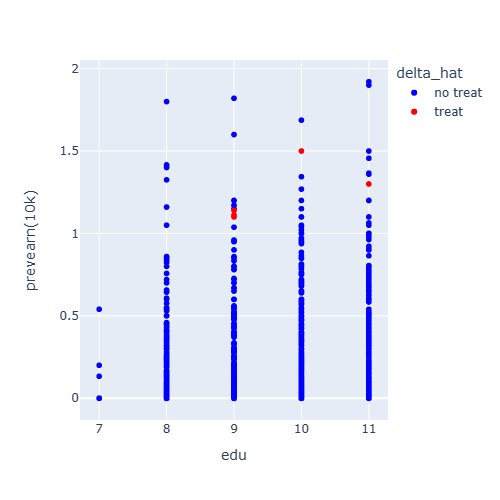}} \subfigure[QoTE with $\tau=0.5$]{
\label{Fig.sub.2} \includegraphics[width=3.9cm,height=3.9cm]{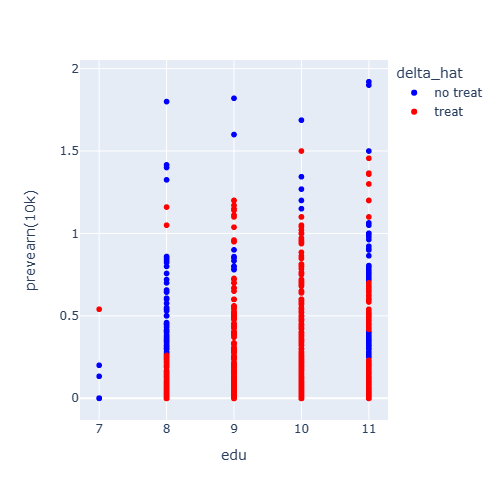}}
\subfigure[QoTE with $\tau=0.75$]{ \label{Fig.sub.3} \includegraphics[width=3.9cm,height=3.9cm]{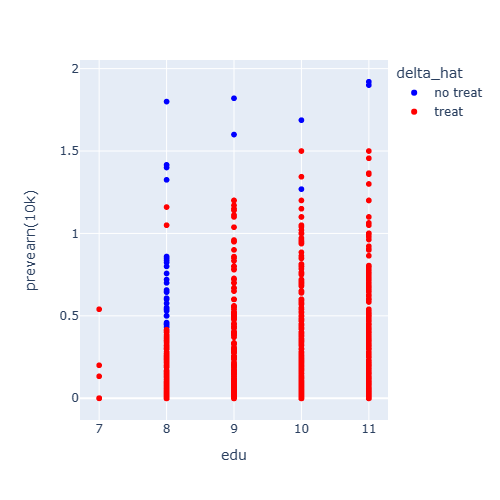}}
\caption{Treatment Decisions for Female Workers Without High School Diploma}
\label{fig:application2-2}
\end{figure}
\begin{figure}[htbp]
\centering \subfigure[QTE with $\tau=0.25$]{ \label{Fig2.sub.1}
\includegraphics[width=3.9cm,height=3.9cm]{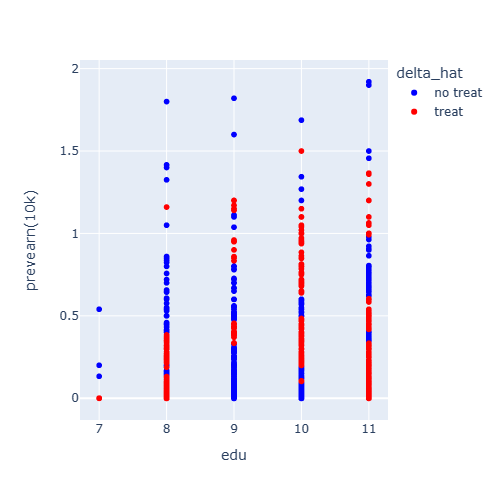}} \subfigure[QTE with $\tau=0.5$]{
\label{Fig2.sub.2} \includegraphics[width=3.9cm,height=3.9cm]{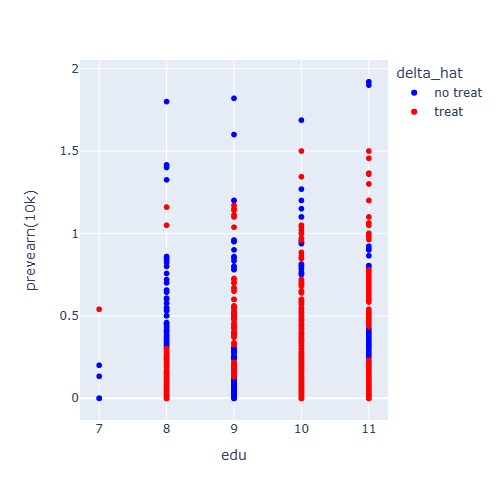}}
\subfigure[QTE with $\tau=0.75$]{ \label{Fig2.sub.3} \includegraphics[width=3.9cm,height=3.9cm]{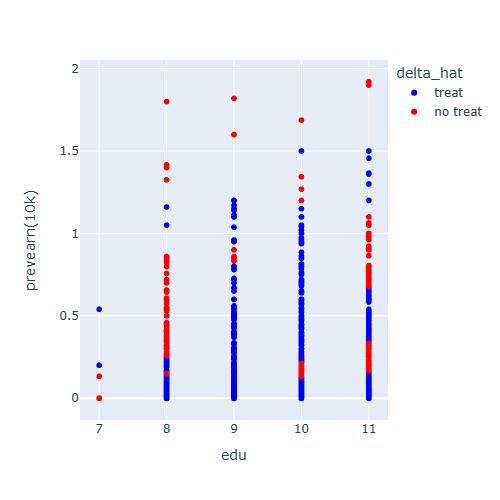}}
\subfigure[ATE]{ \label{Fig2.sub.4} \includegraphics[width=3.9cm,height=3.9cm]{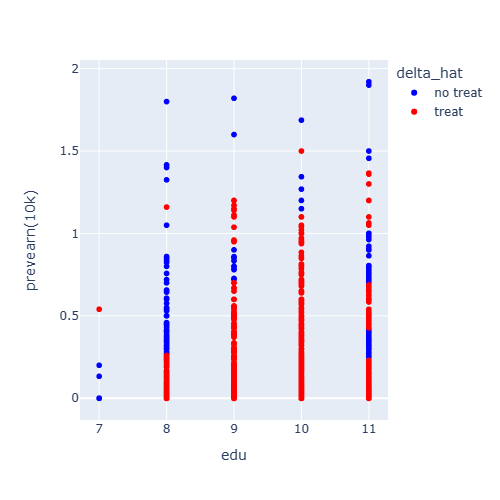}}
\caption{Treatment Decisions Based on the QTEs and the ATE}
\label{fig:application2-3}
\end{figure}

Next, in Figure \ref{fig:application2-2}, we present the decisions
of allocating the job training to the female group without a high
school diploma in the space of education and previous earnings (i.e.,
the two important covariates for the allocation decision). Again,
we use the 0.25-quantile, median, and 0.75-quantile QoTE bounds to
represent prudent, majority-minded, and negligent PMs, respectively.
The 0.75-quantile bounds suggest the treatment option more often than
the other cases. It would be compelling to treat workers suggested
by the 0.25-quantile bounds, as they produce prudent decisions.

For comparison, in Figure \ref{fig:application2-3}, we present the
decisions based on the 0.25-quantile, median, and 0.75-quantile QTE
and the ATE. Again, there is no obvious tendency in decisions when
the quantile probability increases from 0.25 to 0.75, which reflects
the possible limitation of using the QTE as the basis for decisions.
The decisions based on the ATE (i.e., the most common approach in
the literature) look very similar to the decisions based on the median
QoTE bounds, which suggests that the issue of outliers is not serious
in this application. In this sense, the policy based on the median
QoTE bounds can be viewed as a robustness check for the policy based
on the ATE (e.g., \citet{kitagawa2018should}).

\section{Full Numerical Illustrations\label{sec:Numerical-Illustrations}}

The question we want to answer via numerical exercises is: how the
performance of treatment allocations differ across welfare criteria,
especially when the QoTE is partially identified. To facilitate illustration,
we focus on in the case of unconstrained $\mathcal{D}$ and no $X$.

\subsection{Simulation Design\label{subsec:Simulation-Design}}

We consider the following data-generating processes (DGPs). We draw
either $(Y_{1},Y_{0})$ or $(\log Y_{1},\log Y_{0})$ from $N(\mu,\Sigma)$,
where $\mu=(\mu_{1},\mu_{0})'$ and $\Sigma=\bigl(\begin{smallmatrix}\sigma_{1} & \rho_{10}\sqrt{\sigma_{0}\sigma_{1}}\\
\rho_{10}\sqrt{\sigma_{0}\sigma_{1}} & \sigma_{0}
\end{smallmatrix}\bigr)$, and $D$ independently from $Bernoulli(0.5)$. Then, the observed
outcome is generated by $Y=DY_{1}+(1-D)Y_{0}$. Note that, under the
bivariate normal distribution, $Y_{1}|Y_{0}\sim N(\mu_{1}+\rho_{10}\sigma_{1}Z_{0},(1-\rho^{2}\sigma_{1}))$
where $Z_{0}=\frac{Y_{0}-\mu_{0}}{\sigma_{0}}$. Therefore, $Y_{1}$
and $Y_{0}$ satisfying stochastically increasingness (SI) (i.e.,
Assumption \ref{as:PD}(iii)) when $\rho_{10}\ge0$. In fact, this
is also true when $Y_{1}$ and $Y_{0}$ are bivariate log-normal;
they are stochastically increasing when $\rho_{10}\ge0$. When $\mathcal{D}$
is unrestricted, the true optimal policies based on the QoTE, QTE
and ATE can be written as follows:
\begin{align*}
 & \delta^{*}=1\{Q_{\tau}(Y_{1}-Y_{0})>0\}\quad\text{where }Q_{\tau}(Y_{1}-Y_{0})=\mu_{1}-\mu_{0}+\Phi^{-1}(\tau)\sqrt{\sigma_{1}^{2}+\sigma_{0}^{2}-2\rho_{10}\sigma_{1}\sigma_{0}},\\
 & \delta_{QTE}^{*}=1\{Q_{\tau}(Y_{1})-Q_{\tau}(Y_{0})>0\}\quad\text{where }Q_{\tau}(Y_{1})-Q_{\tau}(Y_{0})=\mu_{1}-\mu_{0}+\Phi^{-1}(\tau)(\sigma_{1}-\sigma_{0}),\\
 & \delta_{ATE}^{*}=1\{E[Y_{1}-Y_{0}]>0\}\quad\text{where }E[Y_{1}-Y_{0}]=\mu_{1}-\mu_{0}.
\end{align*}
Note that these policies are first-best regardless of whether we consider
a deterministic or stochastic policy. Unlike $\delta_{QTE}^{*}$ and
$\delta_{ATE}^{*}$, the proposed $\delta^{*}$ involves model uncertainty.
Therefore we consider $\delta_{mmr}^{*}$ that minimizes (4.1). Its
expression for the optimal deterministic and stochastic policies $\delta^{*,determ}$
and $\delta^{*,stoch}$ are given in Section 4.1.

In simulation, the bounds $Q_{\tau}^{L}$ and $Q_{\tau}^{U}$ are
calculated under either no assumption (i.e., Makarov bounds) or SI.
Under the latter, we use discretization to calculate the bounds. For
the population policies $\delta_{mmr}^{*}$, $\delta_{QTE}^{*}$ and
$\delta_{ATE}^{*}$, we estimate their sample counterparts $\hat{\delta}^{*}$,
$\hat{\delta}_{QTE}^{*}$ and $\hat{\delta}_{ATE}^{*}$ by estimating
$Q_{\tau}^{U}$, $Q_{\tau}^{L}$, $Q_{\tau}(Y_{d})$, and $E[Y_{d}]$
($d=0,1$). Since $D$ is exogenous in our simulated data, $Q_{\tau}(Y_{d})=Q_{\tau}(Y|D=d)$
and $E[Y_{d}]=E[Y|D=d]$. For each estimate $\hat{\delta}_{j}$ of
$\delta_{j}^{*}$ ($j\in\{\emptyset,QTE,ATE\}$), the misclassification
error is $\mathbb{E}_{n}[1\{\hat{\delta}_{j}\neq\delta_{j}^{*}\}]$
and regret is defined as $\mathbb{E}_{n}[|T_{j}|\cdot1\{\hat{\delta}_{j}\neq\delta_{j}^{*}\}]$
where $T=Q_{\tau}(Y_{1}-Y_{0})$, $T_{QTE}=Q_{\tau}(Y_{1})-Q_{\tau}(Y_{0})$,
and $T_{ATE}=E[Y_{1}-Y_{0}]$ are the corresponding treatment effects
(or equivalently the welfare gains). We focus on $\tau=0.25$.

\subsection{Simulation Results\label{subsec:Simulation-Results}}

Tables \ref{tab:classification_rate1}--\ref{tab:classification_rate2}
present the simulated correct classification rates of the estimated
policies relative to the (true) population policies. We set $n=1000$
for Table \ref{tab:classification_rate1} and $n=50$ for Table \ref{tab:classification_rate2}.
To calculate each classification rate, we replicate each experiment
$200$ times. We consider both the correct specification of SI and
misspecification. We also vary the parameter values in the normal
and log-normal distributions. We treat each DGP as a subgroup of population
(as if it corresponds to a particular value of $X$ if covariates
were to be introduced). Subgroups 1--4 and 7 follow the normal distribution
and SI, and Subgroups 5--6 are where SI is violated. Under bivariate
normality and SI, if $0<\tau<0.5$, $Q_{\tau}(Y_{1}-Y_{0})<Q_{\tau}(Y_{1})-Q_{\tau}(Y_{0})$
and $Q_{\tau}(Y_{1}-Y_{0})<E(Y_{1})-E(Y_{0})$. The purpose of Subgroup
8 is to break this mechanical relationship. Subgroup 8 follows the
log-normal distribution and SI.

\begin{table}[ht]
\begin{centering}
\begin{tabular}{|c|cccccc|}
\hline 
\diagbox{Optimal Policy}{Estimated Policy} & $\hat{\delta}^{stoch,SI}$ & $\hat{\delta}^{stoch}$ & $\hat{\delta}^{determ,SI}$ & $\hat{\delta}^{determ}$ & $\hat{\delta}_{QTE}$ & $\hat{\delta}_{ATE}$\tabularnewline
\hline 
\hline 
 & \multicolumn{6}{c|}{Subgroup 1}\tabularnewline
\hline 
$\delta^{*}$ & $100\%$ & $100\%$ & $100\%$ & $100\%$ & $1.5\%$ & $100\%$\tabularnewline
$\delta_{QTE}^{*}$ & $0\%$ & $0\%$ & $0\%$ & $0\%$ & $98.5\%$ & $0\%$\tabularnewline
$\delta_{ATE}^{*}$ & $100\%$ & $100\%$ & $100\%$ & $100\%$ & $1.5\%$ & $100\%$\tabularnewline
\hline 
 & \multicolumn{6}{c|}{Subgroup 2}\tabularnewline
\hline 
$\delta^{*}$ & $100\%$ & $99\%$ & $100\%$ & $100\%$ & $0\%$ & $0\%$\tabularnewline
$\delta_{QTE}^{*}$ & $0\%$ & $1\%$ & $0\%$ & $0\%$ & $100\%$ & $100\%$\tabularnewline
$\delta_{ATE}^{*}$ & $0\%$ & $1\%$ & $0\%$ & $0\%$ & $100\%$ & $100\%$\tabularnewline
\hline 
 & \multicolumn{6}{c|}{Subgroup 3}\tabularnewline
\hline 
$\delta^{*}$ & $89\%$ & $59\%$ & $100\%$ & $95\%$ & $100\%$ & $100\%$\tabularnewline
$\delta_{QTE}^{*}$ & $89\%$ & $59\%$ & $100\%$ & $95\%$ & $100\%$ & $100\%$\tabularnewline
$\delta_{ATE}^{*}$ & $89\%$ & $59\%$ & $100\%$ & $95\%$ & $100\%$ & $100\%$\tabularnewline
\hline 
 & \multicolumn{6}{c|}{Subgroup 4}\tabularnewline
\hline 
$\delta^{*}$ & $21\%$ & $43\%$ & $0\%$ & $20\%$ & $0\%$ & $0\%$\tabularnewline
$\delta_{QTE}^{*}$ & $79\%$ & $57\%$ & $100\%$ & $80\%$ & $100\%$ & $100\%$\tabularnewline
$\delta_{ATE}^{*}$ & $79\%$ & $57\%$ & $100\%$ & $80\%$ & $100\%$ & $100\%$\tabularnewline
\hline 
 & \multicolumn{6}{c|}{Subgroup 5}\tabularnewline
\hline 
$\delta^{*}$ & $92\%$ & $74\%$ & $100\%$ & $100\%$ & $100\%$ & $0\%$\tabularnewline
$\delta_{QTE}^{*}$ & $92\%$ & $74\%$ & $100\%$ & $100\%$ & $100\%$ & $0\%$\tabularnewline
$\delta_{ATE}^{*}$ & $8\%$ & $26\%$ & $0\%$ & $0\%$ & $0\%$ & $100\%$\tabularnewline
\hline 
 & \multicolumn{6}{c|}{Subgroup 6}\tabularnewline
\hline 
$\delta^{*}$ & $34\%$ & $48\%$ & $6.5\%$ & $86\%$ & $0\%$ & $0\%$\tabularnewline
$\delta_{QTE}^{*}$ & $66\%$ & $52\%$ & $93.5\%$ & $14\%$ & $100\%$ & $100\%$\tabularnewline
$\delta_{ATE}^{*}$ & $66\%$ & $52\%$ & $93.5\%$ & $14\%$ & $100\%$ & $100\%$\tabularnewline
\hline 
 & \multicolumn{6}{c|}{Subgroup 7}\tabularnewline
\hline 
$\delta^{*}$ & $76.5\%$ & $53\%$ & $100\%$ & $31.5\%$ & $100\%$ & $100\%$\tabularnewline
$\delta_{QTE}^{*}$ & $76.5\%$ & $53\%$ & $100\%$ & $31.5\%$ & $100\%$ & $100\%$\tabularnewline
$\delta_{ATE}^{*}$ & $76.5\%$ & $53\%$ & $100\%$ & $31.5\%$ & $100\%$ & $100\%$\tabularnewline
\hline 
 & \multicolumn{6}{c|}{Subgroup 8 (log normal)}\tabularnewline
\hline 
$\delta^{*}$ & $99.5\%$ & $48\%$ & $100\%$ & $32.5\%$ & $100\%$ & $4.5\%$\tabularnewline
$\delta_{QTE}^{*}$ & $99.5\%$ & $48\%$ & $100\%$ & $32.5\%$ & $100\%$ & $4.5\%$\tabularnewline
$\delta_{ATE}^{*}$ & $0\%$ & $52\%$ & $0\%$ & $67.5\%$ & $0\%$ & $95.5\%$\tabularnewline
\hline 
\end{tabular}
\par\end{centering}
\caption{Correct Classification Rate ($\tau=0.25$, $n=1000$)}
\label{tab:classification_rate1}
\end{table}

\clearpage{}

\begin{table}[ht]
\begin{centering}
\begin{tabular}{|c|cccccc|}
\hline 
\diagbox{Optimal Policy}{Estimated Policy} & $\hat{\delta}^{stoch,SI}$ & $\hat{\delta}^{stoch}$ & $\hat{\delta}^{determ,SI}$ & $\hat{\delta}^{determ}$ & $\hat{\delta}_{QTE}$ & $\hat{\delta}_{ATE}$\tabularnewline
\hline 
\hline 
 & \multicolumn{6}{c|}{Subgroup 1}\tabularnewline
\hline 
$\delta^{*}$ & $100\%$ & $100\%$ & $100\%$ & $100\%$ & $33.5\%$ & $93\%$\tabularnewline
$\delta_{QTE}^{*}$ & $0\%$ & $0\%$ & $0\%$ & $0\%$ & $66.5\%$ & $7\%$\tabularnewline
$\delta_{ATE}^{*}$ & $100\%$ & $100\%$ & $100\%$ & $100\%$ & $33.5\%$ & $93\%$\tabularnewline
\hline 
 & \multicolumn{6}{c|}{Subgroup 2}\tabularnewline
\hline 
$\delta^{*}$ & $92\%$ & $90\%$ & $90.5\%$ & $94\%$ & $1\%$ & $17\%$\tabularnewline
$\delta_{QTE}^{*}$ & $8\%$ & $10\%$ & $9.5\%$ & $6\%$ & $99\%$ & $83\%$\tabularnewline
$\delta_{ATE}^{*}$ & $8\%$ & $10\%$ & $9.5\%$ & $6\%$ & $99\%$ & $83\%$\tabularnewline
\hline 
 & \multicolumn{6}{c|}{Subgroup 3}\tabularnewline
\hline 
$\delta^{*}$ & $79\%$ & $51\%$ & $84\%$ & $62\%$ & $99.5\%$ & $100\%$\tabularnewline
$\delta_{QTE}^{*}$ & $79\%$ & $51\%$ & $84\%$ & $62\%$ & $99.5\%$ & $100\%$\tabularnewline
$\delta_{ATE}^{*}$ & $79\%$ & $51\%$ & $84\%$ & $62\%$ & $99.5\%$ & $100\%$\tabularnewline
\hline 
 & \multicolumn{6}{c|}{Subgroup 4}\tabularnewline
\hline 
$\delta^{*}$ & $26\%$ & $49.5\%$ & $14.5\%$ & $44\%$ & $1.5\%$ & $0.5\%$\tabularnewline
$\delta_{QTE}^{*}$ & $74\%$ & $50.5\%$ & $85.5\%$ & $56\%$ & $98.5\%$ & $99.5\%$\tabularnewline
$\delta_{ATE}^{*}$ & $74\%$ & $50.5\%$ & $85.5\%$ & $56\%$ & $98.5\%$ & $99.5\%$\tabularnewline
\hline 
 & \multicolumn{6}{c|}{Subgroup 5}\tabularnewline
\hline 
$\delta^{*}$ & $82\%$ & $81\%$ & $83\%$ & $93.5\%$ & $66.5\%$ & $4\%$\tabularnewline
$\delta_{QTE}^{*}$ & $82\%$ & $81\%$ & $83\%$ & $93.5\%$ & $66.5\%$ & $4\%$\tabularnewline
$\delta_{ATE}^{*}$ & $18\%$ & $19\%$ & $17\%$ & $6.5\%$ & $33.5\%$ & $96\%$\tabularnewline
\hline 
 & \multicolumn{6}{c|}{Subgroup 6}\tabularnewline
\hline 
$\delta^{*}$ & $43\%$ & $61.5\%$ & $40\%$ & $61\%$ & $24\%$ & $0\%$\tabularnewline
$\delta_{QTE}^{*}$ & $57\%$ & $38.5\%$ & $60\%$ & $39\%$ & $76\%$ & $100\%$\tabularnewline
$\delta_{ATE}^{*}$ & $57\%$ & $38.5\%$ & $60\%$ & $39\%$ & $76\%$ & $100\%$\tabularnewline
\hline 
 & \multicolumn{6}{c|}{Subgroup 7}\tabularnewline
\hline 
$\delta^{*}$ & $71.5\%$ & $48\%$ & $80\%$ & $49.5\%$ & $96\%$ & $99.5\%$\tabularnewline
$\delta_{QTE}^{*}$ & $71.5\%$ & $48\%$ & $80\%$ & $49.5\%$ & $96\%$ & $99.5\%$\tabularnewline
$\delta_{ATE}^{*}$ & $71.5\%$ & $48\%$ & $80\%$ & $49.5\%$ & $96\%$ & $99.5\%$\tabularnewline
\hline 
 & \multicolumn{6}{c|}{Subgroup 8 (log normal)}\tabularnewline
\hline 
$\delta^{*}$ & $81\%$ & $53\%$ & $85\%$ & $48\%$ & $100\%$ & $58.5\%$\tabularnewline
$\delta_{QTE}^{*}$ & $81\%$ & $53\%$ & $85\%$ & $48\%$ & $100\%$ & $58.5\%$\tabularnewline
$\delta_{ATE}^{*}$ & $19\%$ & $47\%$ & $15\%$ & $52\%$ & $0\%$ & $41.5\%$\tabularnewline
\hline 
\end{tabular}
\par\end{centering}
\caption{Correct Classification Rate ($\tau=0.25$, $n=50$)}
\label{tab:classification_rate2}
\end{table}

\clearpage{}

Here are the summary of the features in the DGP and corresponding
results in Tables \ref{tab:classification_rate1}--\ref{tab:classification_rate2}.
Recall that $\tau=0.25$.

\medskip{}

\noindent Overall, the correct classification rate tends to be high
when the welfare criterion of the estimated policy matches that of
the population policy. \medskip{}

\noindent \emph{Subgroup 1}: Both intervals under SI and no assumption
exclude $0$ and lie relatively far from it; therefore, both $\hat{\delta}$
and $\hat{\delta}^{SI}$ perform well; $\delta_{QTE}^{*}\neq\delta^{*}=\delta_{ATE}^{*}$.
\medskip{}

\noindent \emph{Subgroup 2}: Both intervals under SI and no assumption
exclude 0; $\hat{\delta}^{determ}$ does not perform worse than $\hat{\delta}^{determ,SI}$
for $\delta^{*}$ because $Q_{\tau}^{L,SI}-Q_{\tau}^{L}>Q_{\tau}^{U}-Q_{\tau}^{U,SI}$;
$\delta^{*}\neq\delta_{QTE}^{*}=\delta_{ATE}^{*}$. \medskip{}

\noindent \emph{Subgroup }3: Both intervals under SI and no assumption
cover 0 (and the same holds for Subgroups 4--7); $\hat{\delta}^{SI}$
performs better than $\hat{\delta}$; $Q_{\tau}^{L,SI}-Q_{\tau}^{L}>Q_{\tau}^{U}-Q_{\tau}^{U,SI}$
and, under the bivariate normal distribution and SI, $Q_{\tau}(Y_{1}-Y_{0})<Q_{\tau}(Y_{1})-Q_{\tau}(Y_{0})$
and $Q_{\tau}(Y_{1}-Y_{0})<E(Y_{1})-E(Y_{0})$ always hold, and thus
both $\hat{\delta}_{QTE}$ and $\hat{\delta}_{ATE}$ perform well;
$\delta^{*}=\delta_{QTE}^{*}=\delta_{ATE}^{*}=1$. \medskip{}

\noindent \emph{Subgroup }4: Both $\hat{\delta}^{SI}$ and $\hat{\delta}$
perform poorly for $\delta^{*}$ because the bound on $Q_{\tau}(Y_{1}-Y_{0})$
covers zero, and the difference between the upper bound and zero is
larger than the difference between the lower bound and zero; $\delta^{*}\neq\delta_{QTE}^{*}=\delta_{ATE}^{*}$.
\medskip{}

\noindent \emph{Subgroup }5: SI is false but $\hat{\delta}^{SI}$
does not perform so poorly because $Q_{\tau}(Y_{1}-Y_{0})$ is still
covered by a relatively long interval; for the same reason, $\hat{\delta}$
does not perform significantly better; $\delta^{*}=\delta_{QTE}^{*}\neq\delta_{ATE}^{*}$.
\medskip{}

\noindent \emph{Subgroup }6: SI is false and $\hat{\delta}^{SI}$
performs poorly because $Q_{\tau}(Y_{1}-Y_{0})$ is \emph{not} covered
by a relatively long interval; $\hat{\delta}$ does not perform well
because $Q_{\tau}^{L,SI}-Q_{\tau}^{L}<Q_{\tau}^{U}-Q_{\tau}^{U,SI}$;
$\delta^{*}\neq\delta_{QTE}^{*}=\delta_{ATE}^{*}$. \medskip{}

\noindent \emph{Subgroup }7: $\hat{\delta}^{SI}$ makes a correct
decision while $\hat{\delta}$ performs worse; meanwhile, $\hat{\delta}_{QTE}$
and $\hat{\delta}_{ATE}$ perform well.\medskip{}

\noindent \emph{Subgroup }8: The interval under SI excludes $0$ while
the interval under no assumption covers $0$; therefore, $\hat{\delta}^{SI}$
performs better than $\hat{\delta}$; under this log-normal setting
and SI, $Q_{\tau}(Y_{1}-Y_{0})<E(Y_{1})-E(Y_{0})$ may be violated,
which occurs in the current subgroup and thus $\hat{\delta}^{SI}$
performs better than $\hat{\delta}_{ATE}$. 

\subsection{Additional Simulation Results\label{subsec:Additional-Simulation-Results-1}}

Tables \ref{tab:classification_rate1-1} and \ref{tab:classification_rate2-1}
present the classification rates with $\tau=0.75$.
\begin{table}[ht]
\begin{centering}
\begin{tabular}{|c|cccccc|}
\hline 
\diagbox{Optimal Policy}{Estimated Policy} & $\hat{\delta}^{stoch,SI}$ & $\hat{\delta}^{stoch}$ & $\hat{\delta}^{determ,SI}$ & $\hat{\delta}^{determ}$ & $\hat{\delta}_{QTE}$ & $\hat{\delta}_{ATE}$\tabularnewline
\hline 
\hline 
 & \multicolumn{6}{c|}{Subgroup 1}\tabularnewline
\hline 
$\delta^{*}$ & $100\%$ & $100\%$ & $100\%$ & $100\%$ & $1.5\%$ & $100\%$\tabularnewline
$\delta_{QTE}^{*}$ & $0\%$ & $0\%$ & $0\%$ & $0\%$ & $98.5\%$ & $0\%$\tabularnewline
$\delta_{ATE}^{*}$ & $100\%$ & $100\%$ & $100\%$ & $100\%$ & $1.5\%$ & $100\%$\tabularnewline
\hline 
 & \multicolumn{6}{c|}{Subgroup 2}\tabularnewline
\hline 
$\delta^{*}$ & $100\%$ & $99\%$ & $100\%$ & $100\%$ & $0\%$ & $0\%$\tabularnewline
$\delta_{QTE}^{*}$ & $0\%$ & $1\%$ & $0\%$ & $0\%$ & $100\%$ & $100\%$\tabularnewline
$\delta_{ATE}^{*}$ & $0\%$ & $1\%$ & $0\%$ & $0\%$ & $100\%$ & $100\%$\tabularnewline
\hline 
 & \multicolumn{6}{c|}{Subgroup 3}\tabularnewline
\hline 
$\delta^{*}$ & $89\%$ & $59\%$ & $100\%$ & $95\%$ & $100\%$ & $100\%$\tabularnewline
$\delta_{QTE}^{*}$ & $89\%$ & $59\%$ & $100\%$ & $95\%$ & $100\%$ & $100\%$\tabularnewline
$\delta_{ATE}^{*}$ & $89\%$ & $59\%$ & $100\%$ & $95\%$ & $100\%$ & $100\%$\tabularnewline
\hline 
 & \multicolumn{6}{c|}{Subgroup 4}\tabularnewline
\hline 
$\delta^{*}$ & $21\%$ & $43\%$ & $0\%$ & $20\%$ & $0\%$ & $0\%$\tabularnewline
$\delta_{QTE}^{*}$ & $79\%$ & $57\%$ & $100\%$ & $80\%$ & $100\%$ & $100\%$\tabularnewline
$\delta_{ATE}^{*}$ & $79\%$ & $57\%$ & $100\%$ & $80\%$ & $100\%$ & $100\%$\tabularnewline
\hline 
 & \multicolumn{6}{c|}{Subgroup 5}\tabularnewline
\hline 
$\delta^{*}$ & $92\%$ & $74\%$ & $100\%$ & $100\%$ & $100\%$ & $0\%$\tabularnewline
$\delta_{QTE}^{*}$ & $92\%$ & $74\%$ & $100\%$ & $100\%$ & $100\%$ & $0\%$\tabularnewline
$\delta_{ATE}^{*}$ & $8\%$ & $26\%$ & $0\%$ & $0\%$ & $0\%$ & $100\%$\tabularnewline
\hline 
 & \multicolumn{6}{c|}{Subgroup 6}\tabularnewline
\hline 
$\delta^{*}$ & $34\%$ & $48\%$ & $6.5\%$ & $86\%$ & $0\%$ & $0\%$\tabularnewline
$\delta_{QTE}^{*}$ & $66\%$ & $52\%$ & $93.5\%$ & $14\%$ & $100\%$ & $100\%$\tabularnewline
$\delta_{ATE}^{*}$ & $66\%$ & $52\%$ & $93.5\%$ & $14\%$ & $100\%$ & $100\%$\tabularnewline
\hline 
 & \multicolumn{6}{c|}{Subgroup 7}\tabularnewline
\hline 
$\delta^{*}$ & $76.5\%$ & $53\%$ & $100\%$ & $31.5\%$ & $100\%$ & $100\%$\tabularnewline
$\delta_{QTE}^{*}$ & $76.5\%$ & $53\%$ & $100\%$ & $31.5\%$ & $100\%$ & $100\%$\tabularnewline
$\delta_{ATE}^{*}$ & $76.5\%$ & $53\%$ & $100\%$ & $31.5\%$ & $100\%$ & $100\%$\tabularnewline
\hline 
 & \multicolumn{6}{c|}{Subgroup 8 (log normal)}\tabularnewline
\hline 
$\delta^{*}$ & $99.5\%$ & $48\%$ & $100\%$ & $32.5\%$ & $100\%$ & $4.5\%$\tabularnewline
$\delta_{QTE}^{*}$ & $99.5\%$ & $48\%$ & $100\%$ & $32.5\%$ & $100\%$ & $4.5\%$\tabularnewline
$\delta_{ATE}^{*}$ & $0\%$ & $52\%$ & $0\%$ & $67.5\%$ & $0\%$ & $95.5\%$\tabularnewline
\hline 
\end{tabular}
\par\end{centering}
\caption{Correct Classification Rate ($\tau=0.75$, $n=1000$)}
\label{tab:classification_rate1-1}
\end{table}
\begin{table}[ht]
\begin{centering}
\begin{tabular}{|c|cccccc|}
\hline 
\diagbox{Optimal Policy}{Estimated Policy} & $\hat{\delta}^{stoch,SI}$ & $\hat{\delta}^{stoch}$ & $\hat{\delta}^{determ,SI}$ & $\hat{\delta}^{determ}$ & $\hat{\delta}_{QTE}$ & $\hat{\delta}_{ATE}$\tabularnewline
\hline 
\hline 
 & \multicolumn{6}{c|}{Subgroup 1}\tabularnewline
\hline 
$\delta^{*}$ & $100\%$ & $100\%$ & $100\%$ & $100\%$ & $33.5\%$ & $93\%$\tabularnewline
$\delta_{QTE}^{*}$ & $0\%$ & $0\%$ & $0\%$ & $0\%$ & $66.5\%$ & $7\%$\tabularnewline
$\delta_{ATE}^{*}$ & $100\%$ & $100\%$ & $100\%$ & $100\%$ & $33.5\%$ & $93\%$\tabularnewline
\hline 
 & \multicolumn{6}{c|}{Subgroup 2}\tabularnewline
\hline 
$\delta^{*}$ & $92\%$ & $90\%$ & $90.5\%$ & $94\%$ & $1\%$ & $17\%$\tabularnewline
$\delta_{QTE}^{*}$ & $8\%$ & $10\%$ & $9.5\%$ & $6\%$ & $99\%$ & $83\%$\tabularnewline
$\delta_{ATE}^{*}$ & $8\%$ & $10\%$ & $9.5\%$ & $6\%$ & $99\%$ & $83\%$\tabularnewline
\hline 
 & \multicolumn{6}{c|}{Subgroup 3}\tabularnewline
\hline 
$\delta^{*}$ & $79\%$ & $51\%$ & $84\%$ & $62\%$ & $99.5\%$ & $100\%$\tabularnewline
$\delta_{QTE}^{*}$ & $79\%$ & $51\%$ & $84\%$ & $62\%$ & $99.5\%$ & $100\%$\tabularnewline
$\delta_{ATE}^{*}$ & $79\%$ & $51\%$ & $84\%$ & $62\%$ & $99.5\%$ & $100\%$\tabularnewline
\hline 
 & \multicolumn{6}{c|}{Subgroup 4}\tabularnewline
\hline 
$\delta^{*}$ & $26\%$ & $49.5\%$ & $14.5\%$ & $44\%$ & $1.5\%$ & $0.5\%$\tabularnewline
$\delta_{QTE}^{*}$ & $74\%$ & $50.5\%$ & $85.5\%$ & $56\%$ & $98.5\%$ & $99.5\%$\tabularnewline
$\delta_{ATE}^{*}$ & $74\%$ & $50.5\%$ & $85.5\%$ & $56\%$ & $98.5\%$ & $99.5\%$\tabularnewline
\hline 
 & \multicolumn{6}{c|}{Subgroup 5}\tabularnewline
\hline 
$\delta^{*}$ & $82\%$ & $81\%$ & $83\%$ & $93.5\%$ & $66.5\%$ & $4\%$\tabularnewline
$\delta_{QTE}^{*}$ & $82\%$ & $81\%$ & $83\%$ & $93.5\%$ & $66.5\%$ & $4\%$\tabularnewline
$\delta_{ATE}^{*}$ & $18\%$ & $19\%$ & $17\%$ & $6.5\%$ & $33.5\%$ & $96\%$\tabularnewline
\hline 
 & \multicolumn{6}{c|}{Subgroup 6}\tabularnewline
\hline 
$\delta^{*}$ & $43\%$ & $61.5\%$ & $40\%$ & $61\%$ & $24\%$ & $0\%$\tabularnewline
$\delta_{QTE}^{*}$ & $57\%$ & $38.5\%$ & $60\%$ & $39\%$ & $76\%$ & $100\%$\tabularnewline
$\delta_{ATE}^{*}$ & $57\%$ & $38.5\%$ & $60\%$ & $39\%$ & $76\%$ & $100\%$\tabularnewline
\hline 
 & \multicolumn{6}{c|}{Subgroup 7}\tabularnewline
\hline 
$\delta^{*}$ & $71.5\%$ & $48\%$ & $80\%$ & $49.5\%$ & $96\%$ & $99.5\%$\tabularnewline
$\delta_{QTE}^{*}$ & $71.5\%$ & $48\%$ & $80\%$ & $49.5\%$ & $96\%$ & $99.5\%$\tabularnewline
$\delta_{ATE}^{*}$ & $71.5\%$ & $48\%$ & $80\%$ & $49.5\%$ & $96\%$ & $99.5\%$\tabularnewline
\hline 
 & \multicolumn{6}{c|}{Subgroup 8 (log normal)}\tabularnewline
\hline 
$\delta^{*}$ & $81\%$ & $53\%$ & $85\%$ & $48\%$ & $100\%$ & $58.5\%$\tabularnewline
$\delta_{QTE}^{*}$ & $81\%$ & $53\%$ & $85\%$ & $48\%$ & $100\%$ & $58.5\%$\tabularnewline
$\delta_{ATE}^{*}$ & $19\%$ & $47\%$ & $15\%$ & $52\%$ & $0\%$ & $41.5\%$\tabularnewline
\hline 
\end{tabular}
\par\end{centering}
\caption{Correct Classification Rate ($\tau=0.75$, $n=50$)}
\label{tab:classification_rate2-1}
\end{table}
 The overall patterns are analogous to the case with $\tau=0.25$.

\subsection{Details of the DGPs of Subgroups in Simulation\label{sec:Details-of-the}}

Tables \ref{tab:pop_values} and \ref{tab:pop_values-1} show the
details of the DGPs and related outputs used in the simulation results
of Sections \ref{subsec:Simulation-Results} (with $\tau=0.25$) and
\ref{subsec:Additional-Simulation-Results-1} (with $\tau=0.75$),
respectively.
\begin{table}
\begin{centering}
{\scriptsize{}}%
\begin{tabular}{|c|c|c|c|c|c|c|}
\hline 
Subgroup & $(\mu_{1},\mu_{0})$ & $(\sigma_{1}^{2},\sigma_{0}^{2})$ & $\rho_{10}$ & $\delta^{*}$ & $\delta_{QTE}^{*}$ & $\delta_{ATE}^{*}$\tabularnewline
\hline 
\hline 
1 & $(2,3)$ & $(1,9)$ & $0.5$ & $\begin{array}{c}
0\\
(-2.97)
\end{array}$ & $\begin{array}{c}
1\\
(0.34)
\end{array}$ & $\begin{array}{c}
0\\
(-1)
\end{array}$\tabularnewline
\hline 
2 & $(4,3)$ & $(1,25)$ & $0.5$ & $\begin{array}{c}
0\\
(-2.09)
\end{array}$ & $\begin{array}{c}
1\\
(3.7)
\end{array}$ & $\begin{array}{c}
1\\
(1)
\end{array}$\tabularnewline
\hline 
3 & $(7,3)$ & $(9,25)$ & $0.5$ & $\begin{array}{c}
1\\
(1.1)
\end{array}$ & $\begin{array}{c}
1\\
(5.35)
\end{array}$ & $\begin{array}{c}
1\\
(4)
\end{array}$\tabularnewline
\hline 
4 & $(3,1)$ & $(5,5)$ & $0.1$ & $\begin{array}{c}
0\\
(-0.23)
\end{array}$ & $\begin{array}{c}
1\\
(2)
\end{array}$ & $\begin{array}{c}
1\\
(2)
\end{array}$\tabularnewline
\hline 
5 & $(3,2)$ & $(9,1)$ & $-0.5$ & $\begin{array}{c}
0\\
(-1.43)
\end{array}$ & $\begin{array}{c}
0\\
(-0.35)
\end{array}$ & $\begin{array}{c}
1\\
(1)
\end{array}$\tabularnewline
\hline 
6 & $(3,0)$ & $(25,4)$ & $-0.5$ & $\begin{array}{c}
0\\
(-1.21)
\end{array}$ & $\begin{array}{c}
1\\
(0.98)
\end{array}$ & $\begin{array}{c}
1\\
(3)
\end{array}$\tabularnewline
\hline 
7 & $(2,0)$ & $(8,4)$ & $0.5$ & $\begin{array}{c}
1\\
(0.30)
\end{array}$ & $\begin{array}{c}
1\\
(1.44)
\end{array}$ & $\begin{array}{c}
1\\
(2)
\end{array}$\tabularnewline
\hline 
8 & $(3,0)$ & $(2,8)$ & $0.8$ & $\begin{array}{c}
1\\
(\text{-})
\end{array}$ & $\begin{array}{c}
1\\
(4.28)
\end{array}$ & $\begin{array}{c}
0\\
(-1.5)
\end{array}$\tabularnewline
\hline 
\end{tabular}{\scriptsize\par}
\par\end{centering}
\begin{centering}
\vspace{0.2cm}
\par\end{centering}
\begin{centering}
{\scriptsize{}}%
\begin{tabular}{|c|c|c|c|c|c|c|c|}
\hline 
Subgroup & $Q_{\tau}^{L,SI}$ & $Q_{\tau}^{U,SI}$ & $Q_{\tau}^{L}$ & $Q_{\tau}^{U}$ & $\delta^{*,stoch,SI}$ & $\delta^{*,stoch}$ & $Y_{1}-Y_{0}$\tabularnewline
\hline 
\hline 
1 & $-3.48$ & $-1.84$ & $-5.1$ & $-1.1$ & $100\%$ & $100\%$ & $N(-1,7)$\tabularnewline
\hline 
2 & $-2.8$ & $-1.19$ & $-4.50$ & $-0.17$ & $100\%$ & $100\%$ & $N(1,21)$\tabularnewline
\hline 
3 & $-0.74$ & $3.91$ & $-4.83$ & $5.63$ & $84\%$ & $54\%$ & $N(4,19)$\tabularnewline
\hline 
4 & $-0.68$ & $2.54$ & $-3.1$ & $3.38$ & $21\%$ & $48\%$ & $N(2,9)$\tabularnewline
\hline 
5 & $-1.48$ & $0.16$ & $-3.1$ & $0.9$ & $90\%$ & $77\%$ & $N(1,13)$\tabularnewline
\hline 
6 & $-1.24$ & $1.92$ & $-4.3$ & $3.5$ & $39\%$ & $55\%$ & $N(3,39)$\tabularnewline
\hline 
7 & $-0.86$ & $2.17$ & $-3.37$ & $3.21$ & $72\%$ & $49\%$ & $N(2,12-4\sqrt{2})$\tabularnewline
\hline 
8 & $0.38$ & $6.57$ & $-8.5$ & $7.75$ & $100\%$ & $48\%$ & -\tabularnewline
\hline 
\end{tabular}{\scriptsize\par}
\par\end{centering}
\begin{centering}
{\small \begin{tablenotes}
    \item Subgroup 8 is under log-normal transformation, i.e., $(\log Y_{1},\log Y_{0})\sim N(\mu,\Sigma)$.
	\item In the parentheses of $\delta^{*}$, $\delta^{*}_{QTE}$, and $\delta^{*}_{ATE}$ are the values of $Q_{\tau}(Y_{1}-Y_{0})$, $Q_{\tau}(Y_{1})-Q_{\tau}(Y_{0})$, and $E[Y_{1}-Y_{0}]$, respectively.
\end{tablenotes}}
\par\end{centering}
\caption{DGP and Population Values}
\label{tab:pop_values}
\end{table}
\begin{table}
\begin{centering}
{\scriptsize{}}%
\begin{tabular}{|c|c|c|c|c|c|c|}
\hline 
Subgroup & $(\mu_{1},\mu_{0})$ & $(\sigma_{1}^{2},\sigma_{0}^{2})$ & $\rho_{10}$ & $\delta^{*}$ & $\delta_{QTE}^{*}$ & $\delta_{ATE}^{*}$\tabularnewline
\hline 
1 & $(2,3)$ & $(1,9)$ & $0.5$ & $\begin{array}{c}
1\\
(0.79)
\end{array}$ & $\begin{array}{c}
0\\
(-2.35)
\end{array}$ & $\begin{array}{c}
0\\
(-1)
\end{array}$\tabularnewline
\hline 
2 & $(4,3)$ & $(1,25)$ & $0.5$ & $\begin{array}{c}
1\\
(4.09)
\end{array}$ & $\begin{array}{c}
0\\
(-1.70)
\end{array}$ & $\begin{array}{c}
1\\
(1)
\end{array}$\tabularnewline
\hline 
3 & $(7,3)$ & $(9,25)$ & $0.5$ & $\begin{array}{c}
1\\
(6.94)
\end{array}$ & $\begin{array}{c}
1\\
(2.65)
\end{array}$ & $\begin{array}{c}
1\\
(4)
\end{array}$\tabularnewline
\hline 
4 & $(3,1)$ & $(5,5)$ & $0.1$ & $\begin{array}{c}
1\\
(4.02)
\end{array}$ & $\begin{array}{c}
1\\
(2)
\end{array}$ & $\begin{array}{c}
1\\
(2)
\end{array}$\tabularnewline
\hline 
5 & $(3,2)$ & $(9,1)$ & $-0.5$ & $\begin{array}{c}
1\\
(3.43)
\end{array}$ & $\begin{array}{c}
1\\
(2.35)
\end{array}$ & $\begin{array}{c}
1\\
(1)
\end{array}$\tabularnewline
\hline 
6 & $(3,0)$ & $(25,4)$ & $-0.5$ & $\begin{array}{c}
1\\
(7.21)
\end{array}$ & $\begin{array}{c}
1\\
(5.02)
\end{array}$ & $\begin{array}{c}
1\\
(3)
\end{array}$\tabularnewline
\hline 
7 & $(2,0)$ & $(8,4)$ & $0.5$ & $\begin{array}{c}
1\\
(3.70)
\end{array}$ & $\begin{array}{c}
1\\
(2.56)
\end{array}$ & $\begin{array}{c}
1\\
(2)
\end{array}$\tabularnewline
\hline 
8 & $(3,0)$ & $(2,11)$ & $0.8$ & $\begin{array}{c}
1\\
(\text{-})
\end{array}$ & $\begin{array}{c}
1\\
(1.72)
\end{array}$ & $\begin{array}{c}
0\\
(-1.5)
\end{array}$\tabularnewline
\hline 
\end{tabular}{\scriptsize\par}
\par\end{centering}
\begin{centering}
\vspace{0.2cm}
\par\end{centering}
\begin{centering}
{\scriptsize{}}%
\begin{tabular}{|c|c|c|c|c|c|c|c|}
\hline 
Subgroup & $Q_{\tau}^{L,SI}$ & $Q_{\tau}^{U,SI}$ & $Q_{\tau}^{L}$ & $Q_{\tau}^{U}$ & $\delta^{*,stoch,SI}$ & $\delta^{*,stoch}$ & $Y_{1}-Y_{0}$\tabularnewline
\hline 
1 & $-0.16$ & $1.48$ & $-0.92$ & $3.10$ & $95\%$ & $77\%$ & $N(-1,7)$\tabularnewline
\hline 
2 & $3.19$ & $4.80$ & $2.17$ & $6.50$ & $100\%$ & $100\%$ & $N(1,21)$\tabularnewline
\hline 
3 & $4.08$ & $8.74$ & $2.37$ & $12.83$ & $100\%$ & $100\%$ & $N(4,19)$\tabularnewline
\hline 
4 & $1.46$ & $4.68$ & $0.61$ & $7.10$ & $100\%$ & $100\%$ & $N(2,9)$\tabularnewline
\hline 
5 & $1.84$ & $3.48$ & $1.08$ & $5.10$ & $100\%$ & $100\%$ & $N(1,13)$\tabularnewline
\hline 
6 & $4.08$ & $7.2$5 & $2.50$ & $10.25$ & $100\%$ & $100\%$ & $N(3,39)$\tabularnewline
\hline 
7 & $1.82$ & $4.87$ & $0.79$ & $7.37$ & $100\%$ & $100\%$ & $N(2,12-4\sqrt{2})$\tabularnewline
\hline 
8 & $18.07$ & $27.33$ & $11.50$ & $26.69$ & $100\%$ & $100\%$ & -\tabularnewline
\hline 
\end{tabular}{\scriptsize\par}
\par\end{centering}
\begin{centering}
{\small \begin{tablenotes}
    \item Subgroup 8 is under log-normal transformation, i.e., $(\log Y_{1},\log Y_{0})\sim N(\mu,\Sigma)$.
	\item In the parentheses of $\delta^{*}$, $\delta^{*}_{QTE}$, and $\delta^{*}_{ATE}$ are the values of $Q_{\tau}(Y_{1}-Y_{0})$, $Q_{\tau}(Y_{1})-Q_{\tau}(Y_{0})$, and $E[Y_{1}-Y_{0}]$, respectively.
\end{tablenotes}}
\par\end{centering}
\caption{DGP and Population Values}
\label{tab:pop_values-1}
\end{table}

\clearpage{}

\section{Welfare Criteria with Stochastic Rules\label{sec:Welfare-Criteria-with}}

We present a more rigorous formulation of the welfare criteria in
Section 2 when a stochastic rule is considered. Let $A(x)$ is a r.v.
representing the stochastic rule drawn from Bernoulli with parameter
$\delta(x)\equiv P[A(x)=1|X=x]$. Then, by assuming $A(X)\perp Y_{d}|X$
for any $d$ (and using it in the third equality below), we have 
\begin{align*}
E[A(X)Y_{1}+(1-A(X))Y_{0}] & =E[Y_{0}]+E[A(X)(Y_{1}-Y_{0})]\\
 & =E[Y_{0}]+E[A(X)E[Y_{1}-Y_{0}|A(X),X]]\\
 & =E[Y_{0}]+E[A(X)E[Y_{1}-Y_{0}|X]]\\
 & =E[Y_{0}]+E[E[Y_{1}-Y_{0}|X]E[A(X)|X]]\\
 & =E[Y_{0}]+E[E[Y_{1}-Y_{0}|X]\delta(X)]\\
 & =E[\delta(X)Y_{1}+(1-\delta(X))Y_{0}].
\end{align*}
Similarly, motivated from the third line above 
\begin{align*}
E[A(X)Q(Y_{1}-Y_{0}|X)] & =E[Q(Y_{1}-Y_{0}|X)E[A(X)|X]]\\
 & =E[Q(Y_{1}-Y_{0}|X)\delta(X)].
\end{align*}
Based on these results, it suffices to use $\delta(x)$ for \emph{both}
deterministic and stochastic rules in Sections 4 and 6.

\section{Proofs\label{sec:Proofs}}

\subsection{Proof of Lemma 3.1}

\begin{proof}
   For $y<y'$ we want to show that $P(Y_{1}>y_{1}|Y_{0}=y)\le P(Y_{1}>y_{1}|Y_{0}=y')$
(i.e., Assumption PD(iii)), suppressing $X$ for simplicity. For any $u\in g^{-}(0,y)\equiv\{u\in\mathbb{R}:g(0,u)=y\}$
and any $u'\in g^{-}(0,y')\equiv\{u\in\mathbb{R}:g(0,u)=y'\}$, we
have $u<u'$, and thus we have $g(1,u)\le g(1,u')$. Then
\begin{align*}
P[Y_{1}>y_{1}|Y_{0}=y] & =\frac{1}{P[U\in g^{-}(0,y)]}\int_{g^{-}(0,y)}P[Y_{1}>y_{1}|U=u]dF(u)\\
 & =\frac{1}{P[U\in g^{-}(0,y)]}\int_{g^{-}(0,y)}P[g(1,U)>y_{1}|U=u]dF(u)\\
 & \le\frac{1}{P[U\in g^{-}(0,y')]}\int_{g^{-}(0,y')}P[g(1,U)>y_{1}|U=u']dF(u')\\
 & =P[Y_{1}>y_{1}|Y_{0}=y'],
\end{align*}
where the inequality holds by the following argument: First, note
that, for any $u\in\mathbb{R}$, $P[g(1,U)>y_{1}|U=u]$ is either
$1$ or $0$. (1) If $P[g(1,U)>y_{1}|U=u']=1$ for all $u'\in g^{-}(0,y')$
then the weak inequality trivially holds as $P[Y_{1}>y_{1}|Y_{0}=y']=1$.
(2) If $P[g(1,U)>y_{1}|U=u']=0$ for some $u'\in g^{-}(0,y')$, then
$g(1,u)\le g(1,u')$ for all $u\in g^{-}(0,y)$, and thus $P[g(1,U)>y_{1}|U=u]=0$
for all $u\in g^{-}(0,y)$, so the weak inequality holds. 
\end{proof}

\subsection{Proof of Lemma 4.1}
\begin{proof}
Fix $x$ and let $\bar{R}(\delta;x)\equiv\max_{Q_{\tau}(x)\in[Q_{\tau}^{L}(x),Q_{\tau}^{U}(x)]}|Q_{\tau}(x)|1\{A(x)\neq sign(Q_{\tau}(x))\}$.
If $0\le Q_{\tau}^{L}(x)\le Q_{\tau}^{U}(x)$, 
\begin{align*}
\bar{R}(\delta;x) & =|Q_{\tau}^{U}(x)|1\{A(x)\neq1\}=Q_{\tau}^{U}(x)1\{A(x)\neq1\}
\end{align*}
and if $0\ge Q_{\tau}^{U}(x)\ge Q_{\tau}^{L}(x)$, 
\begin{align*}
\bar{R}(\delta;x) & =|Q_{\tau}^{L}(x)|1\{A(x)\neq0\}=-Q_{\tau}^{L}(x)1\{A(x)\neq0\}.
\end{align*}
Finally, if $Q_{\tau}^{L}(x)<0<Q_{\tau}^{U}(x)$, 
\begin{align*}
\bar{R}(\delta;x) & =|Q_{\tau}^{U}(x)|1\{A(x)\neq1\}+|Q_{\tau}^{L}(x)|1\{A(x)\neq0\}\\
 & =Q_{\tau}^{U}(x)1\{A(x)\neq1\}-Q_{\tau}^{L}(x)1\{A(x)\neq0\}.
\end{align*}
Therefore, by the law of iterated expectation and Assumption RC, we
have 
\begin{align}
\bar{R}(\delta) & =E\left[Q_{\tau}^{U}(X)[1-\delta(X)]1\{Q_{\tau}^{L}(X)\ge0\}-Q_{\tau}^{L}(X)\delta(X)1\{Q_{\tau}^{U}(X)\le0\}\right.\nonumber \\
 & +Q_{\tau}^{U}(X)[1-\delta(X)]1\{Q_{\tau}^{L}(X)<0<Q_{\tau}^{U}(X)\}-Q_{\tau}^{L}(X)\delta(X)1\{Q_{\tau}^{L}(X)<0<Q_{\tau}^{U}(X)\}\bigg].\label{eq:max_regret0}
\end{align}
From \eqref{eq:max_regret0}, it is straightforward to show (4.2)
and (4.3). To show (4.4), note that, if $Q_{\tau}^{L}(x)<0<Q_{\tau}^{U}(x)$,
\begin{align*}
 & Q_{\tau}^{U}(x)1\{A(x)\neq1\}-Q_{\tau}^{L}(x)1\{A(x)\neq0\}\\
 & =|Q_{\tau}^{U}(x)+Q_{\tau}^{L}(x)|1\{A(x)\neq sign(Q_{\tau}^{U}(x)+Q_{\tau}^{L}(x))\}+\min(Q_{\tau}^{U}(x),-Q_{\tau}^{L}(x)).
\end{align*}
This can be shown by inspecting each case of $A(x)=1$ and $A(x)=0$.
If $0\le Q_{\tau}^{L}(x)\le Q_{\tau}^{U}(x)$, it is obvious that
$Q_{\tau}^{U}(x)1\{A(x)\neq1\}=Q_{\tau}^{U}(x)1\{A(x)\neq sign(Q_{\tau}^{U}(x))\}$
and similarly for the case of $0\ge Q_{\tau}^{U}(x)\ge Q_{\tau}^{L}(x)$.
Then, by applying the law of iterated expectation, we have the desired
result.
\end{proof}

\subsection{Proof of Theorem 4.1}
\begin{proof}
Given the expression of $\delta^{*,stoch}$, the maximum risk of $\delta^{*,stoch}$
is 
\begin{align*}
\bar{R}(\delta^{*,stoch}) & =E\left[\frac{Q_{\tau}^{L}(X)Q_{\tau}^{U}(X)}{Q_{\tau}^{L}(X)-Q_{\tau}^{U}(X)}1\{Q_{\tau}^{L}(X)<0<Q_{\tau}^{U}(X)\}\right].
\end{align*}
Without loss of generality, suppose $X\in[0,1]^{p}$. Since $R(\delta^{*,stoch})\le\bar{R}(\delta^{*,stoch})$,
we have 
\begin{align*}
R(\hat{\delta}^{*,stoch}) & \leq E\left[\frac{Q_{\tau}^{L}(X)Q_{\tau}^{U}(X)}{Q_{\tau}^{L}(X)-Q_{\tau}^{U}(X)}1\{Q_{\tau}^{L}(X)<0<Q_{\tau}^{U}(X)\}\right]+o_{p}(1),
\end{align*}
because $V(\hat{\delta}^{stoch})-V(\delta^{*,stoch})=o_{p}(1)$, which
can be shown as follows: 
\begin{align*}
V(\hat{\delta}^{stoch})-V(\delta^{*,stoch}) & =E[(\hat{A}(X)-A(X))Q_{\tau}(X)]=o_{p}(1)O(1)=o_{p}(1)
\end{align*}
and $E[\hat{A}(X)-A(X)]=E[\hat{\delta}^{stoch}(X)-\delta^{*,stoch}(X)]=o_{p}(1)$
by the definition of $A(X)$ and the definition of $\hat{A}(X)$ with
the estimated Bernoulli probability $\hat{\delta}^{stoch}(X)$.

Next, given the expression of $\delta^{*,determ}$, the maximum risk
of $\delta^{*,determ}$ is 
\begin{align*}
\bar{R}(\delta^{*,determ}) & =E[\min(\max(Q_{\tau}^{U}(X),0),\max(-Q_{\tau}^{L}(X),0))].
\end{align*}
Again, since $R(\delta^{*,determ})\le\bar{R}(\delta^{*,determ})$
we have 
\begin{align*}
\bar{R}(\hat{\delta}^{determ}) & \leq E[\min(\max(Q_{\tau}^{U}(X),0),\max(-Q_{\tau}^{L}(X),0))]+o_{p}(1),
\end{align*}
because $V(\hat{\delta}^{determ})-V(\delta^{*,determ})=o_{p}(1)$,
which can be shown as follows: 
\begin{align*}
 & V(\hat{\delta}^{determ})-V(\delta^{*,determ})\\
= & E[1(\hat{\delta}^{determ}(X)=\delta^{*,determ}(X))\times0+1(\hat{\delta}^{determ}(X)=1,\delta^{*,determ}(X)=0)\times Q_{\tau}(X)\\
 & -1(\hat{\delta}^{determ}(X)=0,\delta^{*,determ}(X)=1)\times Q_{\tau}(X)]\\
= & 0+o_{p}(1)O(1)=o_{p}(1),
\end{align*}
and $E[1(\hat{\delta}^{determ}(X)\neq\delta^{*,determ}(X))]=P[\hat{\delta}^{determ}(X)\neq\delta^{*,determ}(X)]=o_{p}(1)$
by the definition of $\hat{\delta}^{determ}$ and $\delta^{*,determ}$.
\end{proof}

\subsection{Proof of Theorem 4.2}
\begin{proof}
By Theorem 3.2 of \citet{zhao2012estimating}, we have that 
\begin{align*}
\bar{R}(\hat{f})-\inf_{f}\bar{R}(f) & \leq\bar{R}^{S}(\hat{f})-\inf_{f}\bar{R}^{S}(f).
\end{align*}
We essentially need to bound the right-hand side. Let 
\begin{align*}
\tilde{f}=\arg\min_{f\in\mathcal{H}_{k}} & E[|\bar{Q}_{\tau}(X)|\phi\{sign(\bar{Q}_{\tau}(X))f(X)\}+\lambda_{n}||f||^{2}].
\end{align*}
Note that 
\begin{align*}
 & \bar{R}^{S}(\hat{f})-\inf_{f}\bar{R}^{S}(f)\\
= & \bar{R}^{S}(\hat{f})-\bar{R}^{S}(\tilde{f})+\bar{R}^{S}(\tilde{f})-\inf_{f\in\mathcal{H}_{k}}[\bar{R}^{S}(f)+\lambda_{n}||f||^{2}]+\inf_{f\in\mathcal{H}_{k}}[\bar{R}^{S}(f)+\lambda_{n}||f||^{2}]-\inf_{f}\bar{R}^{S}(f)\\
\leq & \inf_{f\in\mathcal{H}_{k}}[\bar{R}^{S}(f)+\lambda_{n}||f||^{2}]-\inf_{f}\bar{R}^{S}(f)\\
 & -\frac{1}{n}\sum_{i=1}^{n}[|\widehat{\bar{Q}}_{\tau}(X_{i})|\phi\{sign(\hat{Q}_{\tau}(X_{i}))\hat{f}(X_{i})\}+\lambda_{n}||\hat{f}||^{2}-|\widehat{\bar{Q}}_{\tau}(X_{i})|\phi\{sign(\widehat{\bar{Q}}_{\tau}(X_{i}))\tilde{f}(X_{i})\}-\lambda_{n}||\tilde{f}||^{2}]\\
 & +E[|\widehat{\bar{Q}}_{\tau}(X)|\phi\{sign(\hat{Q}_{\tau}(X))\hat{f}(X)\}+\lambda_{n}||\hat{f}||^{2}-|\widehat{\bar{Q}}_{\tau}(X)|\phi\{sign(\widehat{\bar{Q}}_{\tau}(X))\tilde{f}(X)\}-\lambda_{n}||\tilde{f}||^{2}]\\
 & +E[|\widehat{\bar{Q}}_{\tau}(X)|\phi\{sign(\bar{Q}_{\tau}(X))\hat{f}(X)\}]-E[|\widehat{\bar{Q}}_{\tau}(X)|\phi\{sign(\widehat{\bar{Q}}_{\tau}(X))\hat{f}(X)\}]\\
 & +E[|\widehat{\bar{Q}}_{\tau}(X)|\phi\{sign(\hat{Q}_{\tau}(X))\tilde{f}(X)\}]-E[|\bar{Q}_{\tau}(X)|\phi\{sign(\bar{Q}_{\tau}(X))\tilde{f}(X)\}].
\end{align*}
Following the proof of Theorem 1 of \citet{zhao2015doubly}, we have
that 
\begin{align*}
 & \bar{R}^{S}(\hat{f})-\inf_{f}\bar{R}^{S}(f)\\
\leq & a(\lambda_{n})+M_{p}c_{n}^{\frac{2}{p+2}}(n\lambda_{n})^{-\frac{2}{p+2}}+M_{p}\lambda_{n}^{-1/2}c_{n}^{\frac{2}{p+2}}n^{-\frac{2}{p+2}}+K\eta\frac{1}{n\lambda_{n}}+2K\eta\frac{1}{n\lambda_{n}^{1/2}}+O_{p}(n^{-\alpha}\lambda_{n}^{-1/2})
\end{align*}
with probability larger than $1-2\exp(-\eta)$. 
\end{proof}

\subsection{Proof of Lemma 5.1}
\begin{proof}
In terms of notation, let $Q_{\tau}(Y_{d}|X)=F_{d|X}^{-1}(\tau)$.
For any $\epsilon>0$, as $n$ goes to infinity, $P[|\{\hat{F}_{1|X}^{-1}(v)-\hat{F}_{0|X}^{-1}(u)\}-\{F_{1|X}^{-1}(v)-F_{0|X}^{-1}(u)\}|\geq\epsilon]\rightarrow0$.
Therefore, on a set with probability converging to 1, we have for
$F_{1|X}^{-1}(v)-F_{0|X}^{-1}(u)\notin(t-\epsilon,t+\epsilon)$, 
\begin{align*}
\left|\int\int1\{\hat{F}_{1|X}^{-1}(v)-\hat{F}_{0|X}^{-1}(u)\le t\}c(u,v)dudv-\int\int1\{F_{1|X}^{-1}(v)-F_{0|X}^{-1}(u)\le t\}c(u,v)dudv\right| & =0,
\end{align*}
where $c(u,v)$ is the copula density (which is bounded), because
$1\{\hat{F}_{1|X}^{-1}(v)-\hat{F}_{0|X}^{-1}(u)\le t\}=1\{F_{1|X}^{-1}(v)-F_{0|X}^{-1}(u)\le t\}$.
For $F_{1|X}^{-1}(v)-F_{0|X}^{-1}(u)\in(t-\epsilon,t+\epsilon)$,
\begin{align*}
\left|\int\int1\{\hat{F}_{1|X}^{-1}(v)-\hat{F}_{0|X}^{-1}(u)\le t\}c(u,v)dudv-\int\int1\{F_{1|X}^{-1}(v)-F_{0|X}^{-1}(u)\le t\}c(u,v)dudv\right| & \leq O_{p}(\epsilon),
\end{align*}
because $\int\int_{(u,v):F_{1|X}^{-1}(v)-F_{0|X}^{-1}(u)\in(t-\epsilon,t+\epsilon)}c(u,v)dudv=O_{p}(\epsilon)$.
Hence, for the infeasible optimal value $\tilde{F}_{\Delta|X}^{L}(t)$
of the linear program using $\hat{F}_{1|X}^{-1}(v)$ and $\hat{F}_{0|X}^{-1}(u)$,
we have 
\begin{align*}
|\tilde{F}_{\Delta|X}^{L}(t)-F_{\Delta|X}^{L}(t)|=o_{p}(1).
\end{align*}
For the feasible optimal value $\hat{F}_{\Delta}^{L}(t)$ using the
discretization approach, we can show that 
\begin{align*}
\hat{F}_{\Delta}^{L}(t) & =\min_{c(\cdot,\cdot)}\sum_{j=1}^{k}\sum_{i=1}^{k}1\{\hat{F}_{Y_{1}}^{-1}(r(i))-\hat{F}_{Y_{0}}^{-1}(r(j))\leq t\}c(i,j)\rightarrow\tilde{F}_{\Delta}^{L}(t),
\end{align*}
as $k=k(n)$ goes to infinity. Therefore, $|\hat{F}_{\Delta|X}^{L}(t)-F_{\Delta|X}^{L}(t)|=o_{p}(1)$.
We can similarly prove the claim for the upper bound $\hat{F}_{\Delta|X}^{U}(t)$
and bounds that are obtained using the Bernstein approximation.
\end{proof}
\bibliographystyle{ecta}
\bibliography{dist_welfare}

\begin{thebibliography}{83}
\newcommand{\enquote}[1]{``#1''}
\expandafter\ifx\csname natexlab\endcsname\relax\def\natexlab#1{#1}\fi

\bibitem[\protect\citeauthoryear{Abbring and Heckman}{Abbring and Heckman}{2007}]{abbring2007econometric}
\textsc{Abbring, J.~H. and J.~J. Heckman} (2007): \enquote{Econometric evaluation of social programs, part III: Distributional treatment effects, dynamic treatment effects, dynamic discrete choice, and general equilibrium policy evaluation,} \emph{Handbook of Econometrics}, 6, 5145--5303.

\bibitem[\protect\citeauthoryear{Adjaho and Christensen}{Adjaho and Christensen}{2022}]{adjaho2022externally}
\textsc{Adjaho, C. and T.~Christensen} (2022): \enquote{Externally valid treatment choice,} \emph{arXiv preprint arXiv:2205.05561}.

\bibitem[\protect\citeauthoryear{Athey and Wager}{Athey and Wager}{2021}]{athey2021policy}
\textsc{Athey, S. and S.~Wager} (2021): \enquote{Policy learning with observational data,} \emph{Econometrica}, 89, 133--161.

\bibitem[\protect\citeauthoryear{Ben-Michael, Greiner, Imai, and Jiang}{Ben-Michael et~al.}{2021}]{ben2021safe}
\textsc{Ben-Michael, E., D.~J. Greiner, K.~Imai, and Z.~Jiang} (2021): \enquote{Safe policy learning through extrapolation: Application to pre-trial risk assessment,} \emph{arXiv preprint arXiv:2109.11679}.

\bibitem[\protect\citeauthoryear{Bloom, Orr, Bell, Cave, Doolittle, Lin, and Bos}{Bloom et~al.}{1997}]{bloom1997benefits}
\textsc{Bloom, H.~S., L.~L. Orr, S.~H. Bell, G.~Cave, F.~Doolittle, W.~Lin, and J.~M. Bos} (1997): \enquote{The benefits and costs of JTPA Title II-A programs: Key findings from the National Job Training Partnership Act study,} \emph{Journal of Human Resources}, 549--576.

\bibitem[\protect\citeauthoryear{Blundell, Gosling, Ichimura, and Meghir}{Blundell et~al.}{2007}]{blundell2007changes}
\textsc{Blundell, R., A.~Gosling, H.~Ichimura, and C.~Meghir} (2007): \enquote{Changes in the distribution of male and female wages accounting for employment composition using bounds,} \emph{Econometrica}, 75, 323--363.

\bibitem[\protect\citeauthoryear{Chen, Austern, and Syrgkanis}{Chen et~al.}{2023}]{chen2023inference}
\textsc{Chen, Q., M.~Austern, and V.~Syrgkanis} (2023): \enquote{Inference on Optimal Dynamic Policies via Softmax Approximation,} \emph{arXiv preprint arXiv:2303.04416}.

\bibitem[\protect\citeauthoryear{Chernozhukov, Fern{\'a}ndez-Val, Hahn, and Newey}{Chernozhukov et~al.}{2013}]{chernozhukov2013average}
\textsc{Chernozhukov, V., I.~Fern{\'a}ndez-Val, J.~Hahn, and W.~Newey} (2013): \enquote{Average and quantile effects in nonseparable panel models,} \emph{Econometrica}, 81, 535--580.

\bibitem[\protect\citeauthoryear{Chernozhukov, Fern{\'a}ndez-Val, Han, and W{\"u}thrich}{Chernozhukov et~al.}{2024}]{chernozhukov2024estimating}
\textsc{Chernozhukov, V., I.~Fern{\'a}ndez-Val, S.~Han, and K.~W{\"u}thrich} (2024): \enquote{Estimating Causal Effects of Discrete and Continuous Treatments with Binary Instruments,} \emph{arXiv preprint arXiv:2403.05850}.

\bibitem[\protect\citeauthoryear{Chernozhukov and Hansen}{Chernozhukov and Hansen}{2005}]{chernozhukov2005iv}
\textsc{Chernozhukov, V. and C.~Hansen} (2005): \enquote{An IV model of quantile treatment effects,} \emph{Econometrica}, 73, 245--261.

\bibitem[\protect\citeauthoryear{Connors, Speroff, Dawson, Thomas, Harrell, Wagner, Desbiens, Goldman, Wu, Califf et~al.}{Connors et~al.}{1996}]{connors1996effectiveness}
\textsc{Connors, A.~F., T.~Speroff, N.~V. Dawson, C.~Thomas, F.~E. Harrell, D.~Wagner, N.~Desbiens, L.~Goldman, A.~W. Wu, R.~M. Califf, et~al.} (1996): \enquote{The effectiveness of right heart catheterization in the initial care of critically ill patients,} \emph{Journal of the American Medical Association}, 276, 889--897.

\bibitem[\protect\citeauthoryear{Cui}{Cui}{2021}]{cui2021individualized}
\textsc{Cui, Y.} (2021): \enquote{Individualized decision making under partial identification: three perspectives, two optimality results, and one paradox,} \emph{Harvard Data Science Review}, 3, 1--19.

\bibitem[\protect\citeauthoryear{Cui and Tchetgen~Tchetgen}{Cui and Tchetgen~Tchetgen}{2021{\natexlab{a}}}]{cui2021necessary}
\textsc{Cui, Y. and E.~Tchetgen~Tchetgen} (2021{\natexlab{a}}): \enquote{On a necessary and sufficient identification condition of optimal treatment regimes with an instrumental variable,} \emph{Statistics \& Probability Letters}, 178, 109180.

\bibitem[\protect\citeauthoryear{Cui and Tchetgen~Tchetgen}{Cui and Tchetgen~Tchetgen}{2021{\natexlab{b}}}]{cui2021semiparametric}
---\hspace{-.1pt}---\hspace{-.1pt}--- (2021{\natexlab{b}}): \enquote{A semiparametric instrumental variable approach to optimal treatment regimes under endogeneity,} \emph{Journal of the American Statistical Association}, 116, 162--173.

\bibitem[\protect\citeauthoryear{D'Adamo}{D'Adamo}{2021}]{d2021orthogonal}
\textsc{D'Adamo, R.} (2021): \enquote{Orthogonal Policy Learning Under Ambiguity,} \emph{arXiv preprint arXiv:2111.10904}.

\bibitem[\protect\citeauthoryear{Doksum}{Doksum}{1974}]{doksum1974empirical}
\textsc{Doksum, K.} (1974): \enquote{Empirical probability plots and statistical inference for nonlinear models in the two-sample case,} \emph{The Annals of Statistics}, 267--277.

\bibitem[\protect\citeauthoryear{Dud{\'\i}k, Langford, and Li}{Dud{\'\i}k et~al.}{2011}]{dudik2011doubly}
\textsc{Dud{\'\i}k, M., J.~Langford, and L.~Li} (2011): \enquote{Doubly robust policy evaluation and learning,} \emph{arXiv preprint arXiv:1103.4601}.

\bibitem[\protect\citeauthoryear{Fan and Park}{Fan and Park}{2010}]{fan2010sharp}
\textsc{Fan, Y. and S.~S. Park} (2010): \enquote{Sharp bounds on the distribution of treatment effects and their statistical inference,} \emph{Econometric Theory}, 26, 931--951.

\bibitem[\protect\citeauthoryear{Firpo and Ridder}{Firpo and Ridder}{2019}]{firpo2019partial}
\textsc{Firpo, S. and G.~Ridder} (2019): \enquote{Partial identification of the treatment effect distribution and its functionals,} \emph{Journal of Econometrics}, 213, 210--234.

\bibitem[\protect\citeauthoryear{Frandsen and Lefgren}{Frandsen and Lefgren}{2021}]{frandsen2021partial}
\textsc{Frandsen, B.~R. and L.~J. Lefgren} (2021): \enquote{Partial identification of the distribution of treatment effects with an application to the Knowledge is Power Program (KIPP),} \emph{Quantitative Economics}, 12, 143--171.

\bibitem[\protect\citeauthoryear{Han}{Han}{2021}]{han2021comment}
\textsc{Han, S.} (2021): \enquote{Comment: Individualized treatment rules under endogeneity,} \emph{Journal of the American Statistical Association}, 116, 192--195.

\bibitem[\protect\citeauthoryear{Han}{Han}{2023}]{han2023optimal}
---\hspace{-.1pt}---\hspace{-.1pt}--- (2023): \enquote{Optimal dynamic treatment regimes and partial welfare ordering,} \emph{Journal of the American Statistical Association}, 1--11.

\bibitem[\protect\citeauthoryear{Han and Yang}{Han and Yang}{2024}]{han2020sharp}
\textsc{Han, S. and S.~Yang} (2024): \enquote{A Computational Approach to Identification of Treatment Effects for Policy Evaluation,} \emph{Journal of Econometrics}, 240.

\bibitem[\protect\citeauthoryear{Heckman and Honore}{Heckman and Honore}{1990}]{heckman1990empirical}
\textsc{Heckman, J.~J. and B.~E. Honore} (1990): \enquote{The empirical content of the Roy model,} \emph{Econometrica: Journal of the Econometric Society}, 1121--1149.

\bibitem[\protect\citeauthoryear{Heckman, Smith, and Clements}{Heckman et~al.}{1997}]{heckman1997making}
\textsc{Heckman, J.~J., J.~Smith, and N.~Clements} (1997): \enquote{Making the most out of programme evaluations and social experiments: Accounting for heterogeneity in programme impacts,} \emph{The Review of Economic Studies}, 64, 487--535.

\bibitem[\protect\citeauthoryear{Heckman and Smith}{Heckman and Smith}{1998}]{heckman1995assessing}
\textsc{Heckman, J.~J. and J.~A. Smith} (1998): \enquote{Evaluating the Welfare State,} in \emph{Econometrics and Economic Theory in the 20th Century: The Ragnar Frisch Centennial}, ed. by S.~Strom, Cambridge University Press, Econometric Society Monograph Series.

\bibitem[\protect\citeauthoryear{Hirano and Imbens}{Hirano and Imbens}{2001}]{hirano2001estimation}
\textsc{Hirano, K. and G.~W. Imbens} (2001): \enquote{Estimation of causal effects using propensity score weighting: An application to data on right heart catheterization,} \emph{Health Services and Outcomes Research Methodology}, 2, 259--278.

\bibitem[\protect\citeauthoryear{Hirano and Porter}{Hirano and Porter}{2009}]{hirano2009asymptotics}
\textsc{Hirano, K. and J.~R. Porter} (2009): \enquote{Asymptotics for statistical treatment rules,} \emph{Econometrica}, 77, 1683--1701.

\bibitem[\protect\citeauthoryear{Ida, Ishihara, Ito, Kido, Kitagawa, Sakaguchi, and Sasaki}{Ida et~al.}{2022}]{ida2022choosing}
\textsc{Ida, T., T.~Ishihara, K.~Ito, D.~Kido, T.~Kitagawa, S.~Sakaguchi, and S.~Sasaki} (2022): \enquote{Choosing Who Chooses: Selection-driven targeting in energy rebate programs,} \emph{National Bureau of Economic Research}.

\bibitem[\protect\citeauthoryear{Ishihara and Kitagawa}{Ishihara and Kitagawa}{2021}]{ishihara2021evidence}
\textsc{Ishihara, T. and T.~Kitagawa} (2021): \enquote{Evidence aggregation for treatment choice,} \emph{arXiv preprint arXiv:2108.06473}.

\bibitem[\protect\citeauthoryear{Jiang, Song, Li, and Zeng}{Jiang et~al.}{2019}]{jiang2019entropy}
\textsc{Jiang, B., R.~Song, J.~Li, and D.~Zeng} (2019): \enquote{Entropy learning for dynamic treatment regimes,} \emph{Statistica Sinica}, 29, 1633.

\bibitem[\protect\citeauthoryear{Joe}{Joe}{2014}]{joe2014dependence}
\textsc{Joe, H.} (2014): \emph{Dependence modeling with copulas}, CRC press.

\bibitem[\protect\citeauthoryear{Kaji and Cao}{Kaji and Cao}{2023}]{kaji2023assessing}
\textsc{Kaji, T. and J.~Cao} (2023): \enquote{Assessing Heterogeneity of Treatment Effects,} \emph{arXiv preprint arXiv:2306.15048}.

\bibitem[\protect\citeauthoryear{Kallus}{Kallus}{2022}]{kallus2022s}
\textsc{Kallus, N.} (2022): \enquote{What's the Harm? Sharp Bounds on the Fraction Negatively Affected by Treatment,} \emph{Advances in Neural Information Processing Systems}, 35, 15996--16009.

\bibitem[\protect\citeauthoryear{Kallus}{Kallus}{2023}]{kallus2023treatment}
---\hspace{-.1pt}---\hspace{-.1pt}--- (2023): \enquote{Treatment effect risk: Bounds and inference,} \emph{Management Science}, 69, 4579--4590.

\bibitem[\protect\citeauthoryear{Kallus, Mao, and Uehara}{Kallus et~al.}{2021}]{kallus2021causal}
\textsc{Kallus, N., X.~Mao, and M.~Uehara} (2021): \enquote{Causal inference under unmeasured confounding with negative controls: A minimax learning approach,} \emph{arXiv preprint arXiv:2103.14029}.

\bibitem[\protect\citeauthoryear{Kallus and Zhou}{Kallus and Zhou}{2021}]{kallus2021minimax}
\textsc{Kallus, N. and A.~Zhou} (2021): \enquote{Minimax-optimal policy learning under unobserved confounding,} \emph{Management Science}, 67, 2870--2890.

\bibitem[\protect\citeauthoryear{Kasy}{Kasy}{2016}]{kasy2016partial}
\textsc{Kasy, M.} (2016): \enquote{Partial identification, distributional preferences, and the welfare ranking of policies,} \emph{Review of Economics and Statistics}, 98, 111--131.

\bibitem[\protect\citeauthoryear{Kitagawa, Lee, and Qiu}{Kitagawa et~al.}{2023}]{kitagawa2023treatment}
\textsc{Kitagawa, T., S.~Lee, and C.~Qiu} (2023): \enquote{Treatment Choice, Mean Square Regret and Partial Identification,} \emph{arXiv preprint arXiv:2310.06242}.

\bibitem[\protect\citeauthoryear{Kitagawa, Sakaguchi, and Tetenov}{Kitagawa et~al.}{2021}]{kitagawa2021constrained}
\textsc{Kitagawa, T., S.~Sakaguchi, and A.~Tetenov} (2021): \enquote{Constrained classification and policy learning,} \emph{arXiv preprint arXiv:2106.12886}.

\bibitem[\protect\citeauthoryear{Kitagawa and Tetenov}{Kitagawa and Tetenov}{2018}]{kitagawa2018should}
\textsc{Kitagawa, T. and A.~Tetenov} (2018): \enquote{Who should be treated? empirical welfare maximization methods for treatment choice,} \emph{Econometrica}, 86, 591--616.

\bibitem[\protect\citeauthoryear{Kitagawa and Tetenov}{Kitagawa and Tetenov}{2021}]{kitagawa2021equality}
---\hspace{-.1pt}---\hspace{-.1pt}--- (2021): \enquote{Equality-minded treatment choice,} \emph{Journal of Business \& Economic Statistics}, 39, 561--574.

\bibitem[\protect\citeauthoryear{Kock and Preinerstorfer}{Kock and Preinerstorfer}{2024}]{kock2024regularizing}
\textsc{Kock, A.~B. and D.~Preinerstorfer} (2024): \enquote{Regularizing Discrimination in Optimal Policy Learning with Distributional Targets,} \emph{arXiv preprint arXiv:2401.17909}.

\bibitem[\protect\citeauthoryear{Kock, Preinerstorfer, and Veliyev}{Kock et~al.}{2022}]{kock2022functional}
\textsc{Kock, A.~B., D.~Preinerstorfer, and B.~Veliyev} (2022): \enquote{Functional sequential treatment allocation,} \emph{Journal of the American Statistical Association}, 117, 1311--1323.

\bibitem[\protect\citeauthoryear{Kock, Preinerstorfer, and Veliyev}{Kock et~al.}{2023}]{kock2023treatment}
---\hspace{-.1pt}---\hspace{-.1pt}--- (2023): \enquote{Treatment recommendation with distributional targets,} \emph{Journal of Econometrics}, 234, 624--646.

\bibitem[\protect\citeauthoryear{Kosorok and Laber}{Kosorok and Laber}{2019}]{kosorok2019precision}
\textsc{Kosorok, M.~R. and E.~B. Laber} (2019): \enquote{Precision medicine,} \emph{Annual Review of Atatistics and Its Application}, 6, 263--286.

\bibitem[\protect\citeauthoryear{Kosorok and Moodie}{Kosorok and Moodie}{2015}]{kosorok2015adaptive}
\textsc{Kosorok, M.~R. and E.~E. Moodie} (2015): \emph{Adaptive treatment strategies in practice: planning trials and analyzing data for personalized medicine}, SIAM.

\bibitem[\protect\citeauthoryear{Lee and Park}{Lee and Park}{2023}]{lee2023nonparametric}
\textsc{Lee, J.~H. and B.~G. Park} (2023): \enquote{Nonparametric identification and estimation of the extended Roy model,} \emph{Journal of Econometrics}, 235, 1087--1113.

\bibitem[\protect\citeauthoryear{Lee}{Lee}{2024}]{lee2021partial}
\textsc{Lee, S.} (2024): \enquote{Partial identification and inference for conditional distributions of treatment effects,} \emph{Journal of Applied Econometrics}, 39, 107--127.

\bibitem[\protect\citeauthoryear{Lehmann}{Lehmann}{1975}]{lehmann1975statistical}
\textsc{Lehmann, E.~L.} (1975): \enquote{Statistical methods based on ranks,} \emph{Nonparametrics. San Francisco, CA, Holden-Day}.

\bibitem[\protect\citeauthoryear{Leqi and Kennedy}{Leqi and Kennedy}{2021}]{leqi2021median}
\textsc{Leqi, L. and E.~H. Kennedy} (2021): \enquote{Median optimal treatment regimes,} \emph{arXiv preprint arXiv:2103.01802}.

\bibitem[\protect\citeauthoryear{Linn, Laber, and Stefanski}{Linn et~al.}{2017}]{linn2017interactive}
\textsc{Linn, K.~A., E.~B. Laber, and L.~A. Stefanski} (2017): \enquote{Interactive q-learning for quantiles,} \emph{Journal of the American Statistical Association}, 112, 638--649.

\bibitem[\protect\citeauthoryear{Makarov}{Makarov}{1982}]{makarov1982estimates}
\textsc{Makarov, G.} (1982): \enquote{Estimates for the distribution function of a sum of two random variables when the marginal distributions are fixed,} \emph{Theory of Probability \& its Applications}, 26, 803--806.

\bibitem[\protect\citeauthoryear{Manski}{Manski}{2004}]{manski2004statistical}
\textsc{Manski, C.~F.} (2004): \enquote{Statistical treatment rules for heterogeneous populations,} \emph{Econometrica}, 72, 1221--1246.

\bibitem[\protect\citeauthoryear{Manski}{Manski}{2007}]{manski2007minimax}
---\hspace{-.1pt}---\hspace{-.1pt}--- (2007): \enquote{Minimax-regret treatment choice with missing outcome data,} \emph{Journal of Econometrics}, 139, 105--115.

\bibitem[\protect\citeauthoryear{Manski and Tetenov}{Manski and Tetenov}{2023}]{manski2023statistical}
\textsc{Manski, C.~F. and A.~Tetenov} (2023): \enquote{Statistical decision theory respecting stochastic dominance,} \emph{The Japanese Economic Review}, 1--23.

\bibitem[\protect\citeauthoryear{Mbakop and Tabord-Meehan}{Mbakop and Tabord-Meehan}{2021}]{mbakop2021model}
\textsc{Mbakop, E. and M.~Tabord-Meehan} (2021): \enquote{Model selection for treatment choice: Penalized welfare maximization,} \emph{Econometrica}, 89, 825--848.

\bibitem[\protect\citeauthoryear{Murphy}{Murphy}{2003}]{murphy2003optimal}
\textsc{Murphy, S.~A.} (2003): \enquote{Optimal dynamic treatment regimes,} \emph{Journal of the Royal Statistical Society Series B: Statistical Methodology}, 65, 331--355.

\bibitem[\protect\citeauthoryear{Pu and Zhang}{Pu and Zhang}{2021}]{pu2021estimating}
\textsc{Pu, H. and B.~Zhang} (2021): \enquote{Estimating optimal treatment rules with an instrumental variable: A partial identification learning approach,} \emph{Journal of the Royal Statistical Society Series B: Statistical Methodology}, 83, 318--345.

\bibitem[\protect\citeauthoryear{Qi, Miao, and Zhang}{Qi et~al.}{2023{\natexlab{a}}}]{qi2023proximal}
\textsc{Qi, Z., R.~Miao, and X.~Zhang} (2023{\natexlab{a}}): \enquote{Proximal learning for individualized treatment regimes under unmeasured confounding,} \emph{Journal of the American Statistical Association}, 1--14.

\bibitem[\protect\citeauthoryear{Qi, Pang, and Liu}{Qi et~al.}{2023{\natexlab{b}}}]{qi2023robustness}
\textsc{Qi, Z., J.-S. Pang, and Y.~Liu} (2023{\natexlab{b}}): \enquote{On robustness of individualized decision rules,} \emph{Journal of the American Statistical Association}, 118, 2143--2157.

\bibitem[\protect\citeauthoryear{Qian and Murphy}{Qian and Murphy}{2011}]{qian2011performance}
\textsc{Qian, M. and S.~A. Murphy} (2011): \enquote{Performance guarantees for individualized treatment rules,} \emph{The Annals of Statistics}, 39, 1180.

\bibitem[\protect\citeauthoryear{Qiu, Carone, Sadikova, Petukhova, Kessler, and Luedtke}{Qiu et~al.}{2021}]{qiu2021optimal}
\textsc{Qiu, H., M.~Carone, E.~Sadikova, M.~Petukhova, R.~C. Kessler, and A.~Luedtke} (2021): \enquote{Optimal individualized decision rules using instrumental variable methods,} \emph{Journal of the American Statistical Association}, 116, 174--191.

\bibitem[\protect\citeauthoryear{Robins}{Robins}{2004}]{robins2004optimal}
\textsc{Robins, J.~M.} (2004): \enquote{Optimal structural nested models for optimal sequential decisions,} in \emph{Proceedings of the Second Seattle Symposium in Biostatistics: Analysis of Correlated Data}, Springer, 189--326.

\bibitem[\protect\citeauthoryear{Rubin and van~der Laan}{Rubin and van~der Laan}{2012}]{rubin2012statistical}
\textsc{Rubin, D.~B. and M.~J. van~der Laan} (2012): \enquote{Statistical issues and limitations in personalized medicine research with clinical trials,} \emph{The International Journal of Biostatistics}, 8, 18.

\bibitem[\protect\citeauthoryear{Sancetta and Satchell}{Sancetta and Satchell}{2004}]{sancetta2004bernstein}
\textsc{Sancetta, A. and S.~Satchell} (2004): \enquote{The Bernstein copula and its applications to modeling and approximations of multivariate distributions,} \emph{Econometric Theory}, 20, 535--562.

\bibitem[\protect\citeauthoryear{Savage}{Savage}{1951}]{savage1951theory}
\textsc{Savage, L.~J.} (1951): \enquote{The theory of statistical decision,} \emph{Journal of the American Statistical Association}, 46, 55--67.

\bibitem[\protect\citeauthoryear{Shen and Cui}{Shen and Cui}{2023}]{shen2023optimal}
\textsc{Shen, T. and Y.~Cui} (2023): \enquote{Optimal treatment regimes for proximal causal learning,} \emph{NeurIPS}.

\bibitem[\protect\citeauthoryear{Shi, Fan, Song, and Lu}{Shi et~al.}{2018}]{shi2018high}
\textsc{Shi, C., A.~Fan, R.~Song, and W.~Lu} (2018): \enquote{High-dimensional A-learning for optimal dynamic treatment regimes,} \emph{The Annals of Statistics}, 46, 925.

\bibitem[\protect\citeauthoryear{Steinwart and Scovel}{Steinwart and Scovel}{2007}]{steinwart2007fast}
\textsc{Steinwart, I. and C.~Scovel} (2007): \enquote{Fast rates for support vector machines using Gaussian kernels,} \emph{The Annals of Statistics}, 35, 575--607.

\bibitem[\protect\citeauthoryear{Stoye}{Stoye}{2007}]{stoye2007minimax}
\textsc{Stoye, J.} (2007): \enquote{Minimax regret treatment choice with incomplete data and many treatments,} \emph{Econometric Theory}, 23, 190--199.

\bibitem[\protect\citeauthoryear{Stoye}{Stoye}{2009}]{stoye2009minimax}
---\hspace{-.1pt}---\hspace{-.1pt}--- (2009): \enquote{Minimax regret treatment choice with finite samples,} \emph{Journal of Econometrics}, 151, 70--81.

\bibitem[\protect\citeauthoryear{Stoye}{Stoye}{2012}]{stoye2012minimax}
---\hspace{-.1pt}---\hspace{-.1pt}--- (2012): \enquote{Minimax regret treatment choice with covariates or with limited validity of experiments,} \emph{Journal of Econometrics}, 166, 138--156.

\bibitem[\protect\citeauthoryear{Tsiatis, Davidian, Holloway, and Laber}{Tsiatis et~al.}{2019}]{tsiatis2019dynamic}
\textsc{Tsiatis, A.~A., M.~Davidian, S.~T. Holloway, and E.~B. Laber} (2019): \emph{Dynamic treatment regimes: Statistical methods for precision medicine}, CRC press.

\bibitem[\protect\citeauthoryear{Villani}{Villani}{2009}]{villani2009optimal}
\textsc{Villani, C.} (2009): \emph{Optimal transport: old and new}, vol. 338, Springer.

\bibitem[\protect\citeauthoryear{Vuong and Xu}{Vuong and Xu}{2017}]{vuong2017counterfactual}
\textsc{Vuong, Q. and H.~Xu} (2017): \enquote{Counterfactual mapping and individual treatment effects in nonseparable models with binary endogeneity,} \emph{Quantitative Economics}, 8, 589--610.

\bibitem[\protect\citeauthoryear{Wang, Zhou, Song, and Sherwood}{Wang et~al.}{2018}]{wang2018quantile}
\textsc{Wang, L., Y.~Zhou, R.~Song, and B.~Sherwood} (2018): \enquote{Quantile-optimal treatment regimes,} \emph{Journal of the American Statistical Association}, 113, 1243--1254.

\bibitem[\protect\citeauthoryear{Watkins and Dayan}{Watkins and Dayan}{1992}]{watkins1992q}
\textsc{Watkins, C.~J. and P.~Dayan} (1992): \enquote{Q-learning,} \emph{Machine learning}, 8, 279--292.

\bibitem[\protect\citeauthoryear{Williamson and Downs}{Williamson and Downs}{1990}]{williamson1990probabilistic}
\textsc{Williamson, R.~C. and T.~Downs} (1990): \enquote{Probabilistic arithmetic. I. Numerical methods for calculating convolutions and dependency bounds,} \emph{International Journal of Approximate Reasoning}, 4, 89--158.

\bibitem[\protect\citeauthoryear{Yata}{Yata}{2021}]{yata2021optimal}
\textsc{Yata, K.} (2021): \enquote{Optimal decision rules under partial identification,} \emph{arXiv preprint arXiv:2111.04926}.

\bibitem[\protect\citeauthoryear{Zhang, Tsiatis, Laber, and Davidian}{Zhang et~al.}{2012}]{zhang2012robust}
\textsc{Zhang, B., A.~A. Tsiatis, E.~B. Laber, and M.~Davidian} (2012): \enquote{A robust method for estimating optimal treatment regimes,} \emph{Biometrics}, 68, 1010--1018.

\bibitem[\protect\citeauthoryear{Zhao, Zeng, Rush, and Kosorok}{Zhao et~al.}{2012}]{zhao2012estimating}
\textsc{Zhao, Y., D.~Zeng, A.~J. Rush, and M.~R. Kosorok} (2012): \enquote{Estimating individualized treatment rules using outcome weighted learning,} \emph{Journal of the American Statistical Association}, 107, 1106--1118.

\bibitem[\protect\citeauthoryear{Zhao, Zeng, Laber, Song, Yuan, and Kosorok}{Zhao et~al.}{2015}]{zhao2015doubly}
\textsc{Zhao, Y.~Q., D.~Zeng, E.~B. Laber, R.~Song, M.~Yuan, and M.~R. Kosorok} (2015): \enquote{Doubly robust learning for estimating individualized treatment with censored data,} \emph{Biometrika}, 102, 151--168.

\end{thebibliography}

\end{document}